\documentclass[conference,compsoc]{IEEEtran}
\IEEEoverridecommandlockouts

\usepackage{pifont}
\usepackage{cite}
\usepackage{amsmath,amssymb,amsfonts, mathtools}
\usepackage{algorithmic}
\usepackage{graphicx}
\usepackage{textcomp}
\usepackage{xcolor}
\usepackage{lipsum}
\usepackage{booktabs}
\usepackage{multirow}
\usepackage{amsthm}
\usepackage{amsmath}
\usepackage{caption}
\usepackage{subcaption}
\usepackage{varwidth}
\usepackage{tcolorbox}
\usepackage{array}
\usepackage{cleveref}
\usepackage{longtable}
\crefname{section}{§}{§§}
\Crefname{section}{§}{§§}

\newcolumntype{L}[1]{>{\raggedright\let\newline\\\arraybackslash\hspace{0pt}}m{#1}}
\newcolumntype{C}[1]{>{\centering\let\newline\\\arraybackslash\hspace{0pt}}m{#1}}
\newcolumntype{R}[1]{>{\raggedleft\let\newline\\\arraybackslash\hspace{0pt}}m{#1}}

\theoremstyle{definition}

\def\BibTeX{{\rm B\kern-.05em{\sc i\kern-.025em b}\kern-.08em
    T\kern-.1667em\lower.7ex\hbox{E}\kern-.125emX}}
\begin{document}

\title{An investigation of security controls and MITRE ATT\&CK techniques}

\author{\IEEEauthorblockN{Md Rayhanur Rahman, Laurie Williams}
}

\maketitle

\begin{abstract}
Attackers utilize a plethora of adversarial techniques in cyberattacks to compromise the confidentiality, integrity, and availability of the target organizations and systems. Information security standards such as NIST, ISO/IEC specify hundreds of security controls that organizations can enforce to protect and defend the information systems from adversarial techniques. However, implementing all the available controls at the same time can be infeasible and security controls need to be investigated in terms of their mitigation ability over adversarial techniques used in cyberattacks as well. \textit{The goal of this research is to aid organizations in making informed choices on security controls to defend against cyberthreats through an investigation of adversarial techniques used in current cyberattacks.} In this study, we investigated the extent of mitigation of 298 NIST SP800-53 controls over 188 adversarial techniques used in 669 cybercrime groups and malware cataloged in the MITRE ATT\&CK framework based upon an existing mapping between the controls and techniques. We identify that, based on the mapping, only 101 out of 298 control are capable of mitigating adversarial techniques. However, we also identify that 53 adversarial techniques cannot be mitigated by any existing controls, and these techniques primarily aid adversaries in bypassing system defense and discovering targeted system information. We identify a set of 20 critical controls that can mitigate 134 adversarial techniques, and on average, can mitigate 72\% of all techniques used by 98\% of the cataloged adversaries in MITRE ATT\&CK. We urge organizations, that do not have any controls enforced in place, to implement the top controls identified in the study. 

\end{abstract}


\section{Introduction}
Information technology (IT) systems have been gaining continuous attention from threat actors with financial motives~\cite{hackernews-financial-backup} and organized backing~\cite{reuters}. For example, Checkpoint reports a 50\% increase in attacks per week on corporate networks in 2021 compared to 2020~\cite{checkpoint}. Forbes reports that a recent penetration testing project finds cyber-criminals can penetrate 93\% of corporate networks~\cite{forbes, betanews}. Attackers utilize a plethora of tactics, techniques, and procedures (TTPs) in cyberattacks to compromise the confidentiality, integrity, and availability of the targeted systems. Cybersecurity researchers and practitioners identify and design security controls to protect and defend information systems and organizations from adversarial TTPs. \textit{Security controls} are the safeguards and countermeasures to protect the confidentiality, integrity, and availability of information systems by satisfying security requirements defined by the organizations~\cite{control-definition}. The controls specify the mitigation framework and guidelines to avoid, detect, counteract, or minimize security risks on information systems, data, assets, and people. For example, NIST SP800-53~\cite{NIST800r53} specifies a control \textit{AC-7: unsuccessful logon attempts} which requires organizations to enforce a limit of consecutive unsuccessful login attempts. \textit{AC-7} mitigates an adversarial technique \textit{T1110: brute force}~\cite{ttp-t1110} where a user can gain access to systems by guessing the password using consecutive attempts.

Cybersecurity organizations and vendors have developed security control frameworks and standards, such as NIST SP800-53~\cite{NIST800r53}, ISO/IEC 27001~\cite{ISO27001}, and CIS~\cite{cistop20}. Although IT organizations implement, integrate, and enforce security controls in their operations, cyberattacks continue to grow. For example, in the Colonial Pipeline ransomware attack~\cite{colonial-pipeline}, the attackers gained unauthorized access by logging in to a Virtual Private Network (VPN) using an already-compromised password (a violation of NIST 800-53 identification and authentication controls IA-5~\cite{NIST800r53}), as reported in the House Committee on Homeland Security hearing~\cite{mandiant-colonial-pipeline}. Experts express concerns that most critical infrastructure does not enforce mandatory controls~\cite{wa-post, us-gao}. The adoption of security controls is especially lacking in small and medium businesses (SMBs). Security vendor BullGuard discovered that 43\% of SMBs in USA and UK do not have any endpoint security in place~\cite{bullguard-research}. Moreover, a survey in 2022 on 2,000 SMBs in USA reveals that only 5\% of the owners consider cybersecurity as the biggest risk~\cite{cnbc-research}. Overall, these examples substantiate that organizations are not prepared enough to defend against cyberattacks and they need to enforce security controls in their environment. 

To achieve protection against cyberthreats, organizations can implement security controls, as suggested in NIST SP800-53 or ISO/IEC 27001. However, as both NIST and ISO/IEC catalog hundreds of controls, implementing all of them can be infeasible. Organizations that do not have any controls in place can prioritize the implementation of the most important controls as a starting point in terms of mitigation of adversarial techniques. Organizations that have already enforced controls, need to continuously assess the effectiveness of the controls based on their ability to mitigate the more frequent adversarial techniques. In the literature, researchers studied the optimal selection of controls~\cite{hadarCyberDigitalTwin2020, dutta2019and, liDefendingAdvancedPersistent2018, sawikSelectionOptimalCountermeasure2013} based upon budget, assets, and risk. However, controls need to be investigated in terms of their mitigation ability over adversarial techniques used in cyberattacks as well. \textit{The goal of this research is to aid organizations in making informed choices on security controls to defend against cyberthreats through an investigation of adversarial techniques used in current cyberattacks.} We state the following research questions (RQ) below.

\begin{itemize}
    \item \textbf{RQ1}: To what extent do security controls mitigate the adversarial techniques used in cyberattacks? 
    \item \textbf{RQ2}: What controls need be adopted first in terms of the extent of mitigation of adversarial techniques?
\end{itemize}

We investigate the extent of mitigation of 298 NIST SP800-53 controls~\cite{NIST800r53} over 188 adversarial techniques cataloged in the MITRE ATT\&CK~\cite{attack} framework based upon an existing mapping between the controls and techniques~\cite{attack-control-mapping} conducted by Center for Threat Informed Defense (CTID)~\cite{CTID}. We propose a security control metric suite to aid organizations in making an informed choice on security controls. To answer RQ1, we evaluate the metric suite on 298 NIST SP800-53~\cite{NIST800r53} security controls, and 188 adversarial techniques cataloged in the MITRE ATT\&CK~\cite{attack} framework. To answer RQ2, we apply clustering on the controls based on the metrics to identify the set of controls that mitigates the majority of the techniques used in current cyberattacks. We contribute the following:

\begin{enumerate}

    \item A security control metric suite for characterizing to what extent a control mitigates adversarial techniques. The metrics provide a cost-agnostic recommendation on choosing the security controls. Organizations having no controls in place can use the metric to determine the choice of controls. Organizations that have enforced controls can use the metric suite to assess the effectiveness of their enforced controls (see~\cref{sec:metrics});
    
    \item Evaluation of the NIST SP800-53 security controls  on the (i) extent of mitigating ATT\&CK techniques; and (ii) extent of mitigating techniques used in 125 cybercrime groups and 544 malware cataloged in ATT\&CK based on the proposed metric suite(see~\cref{sec:rq1});
    
    \item Identification of 20 critical controls that should be adopted first to achieve protection against the majority of the adversarial techniques(see~\cref{sec:rq2});
    
    \item A set of actionable takeaways on how organizations can use the proposed metrics in practice(see~\cref{takeaways});
    
    \item Dataset and source code for this research made available at~\cite{gitrepo} for replication, including re-running the research when the ATT\&CK Framework and/or the ATT\&CK-NIST mapping are updated. 
    
\end{enumerate}

The rest of the paper is organized as follows.~\cref{sec:metrics} discusses the proposed metric suite.~\cref{method} discusses the methodology.~\cref{sec:rq1},~\cref{sec:rq2}, and~\cref{summary} discuss the findings.~\cref{takeaways}, and~\cref{limitation} discuss the takeaways and limitations followed by related work(~\cref{sec:related-work}) and conclusion(~\cref{sec:conclusion}).

\section{Key Concepts}
This section discusses several related key concepts.
\subsection{Security Controls and NIST SP800-53}
\textit{Security controls} refers to the safeguards and countermeasures to protect the confidentiality, integrity, and availability of information systems by satisfying security requirements defined by the organizations~\cite{control-definition}. By using security controls, organizations can protect hardware, software, networks, and data from malicious actions from adversaries causing loss or damage. The NIST SP800-53 revision 5~\cite{NIST800r53} catalogs 298 security controls across 20  categories. The controls follow the naming convention \textit{CC-i} which denotes the $i^{th}$ control in \textit{CC} category. For example. \textit{CM-2} is the $2^{nd}$ control in the \textit{CM} (configuration management) category.

\subsection{Tactics, Techniques, and Procedures (TTPs)}
Adversaries utilize a variety of tactics, techniques, and procedures (TTPs) to compromise the confidentiality, integrity, and availability of target systems~\cite{attack-design, attack}. \textit{Tactic} refers to the adversaries' overall goal for performing malicious actions. \textit{Techniques} refers to the method adversaries perform to achieve the tactical goal. \textit{Procedure} refers to the specific step-by-step implementation of the corresponding technique. Overall, adversary behavior and stages of an attack can be analyzed and profiles based upon what TTPs are used by the adversaries~\cite{nist-glossary}. For example, in the ATT\&CK framework, adversaries usee \textit{privilege escalation, TA0004}~\cite{ta0004} to gain elevated permission on a system. Adversaries can use \textit{access token manipulation, T1134}~\cite{attack} techniques for escalating the privilege. Thus, adversaries can tamper with access tokens to gain elevated privilege in an environment. A procedure for manipulating access token is \textit{using Metasploit's named-pipe impersonation}~\cite{metasploit}, which has been implemented by the FIN6 cybercrime group~\cite{fin6}. 

\subsection{MITRE ATT\&CK}
The MITRE~\cite{mitre-org} organization developed the ATT\&CK~\cite{attack} framework in 2013, which catalogs of TTPs based on real-world observations of cyberattacks. ATT\&CK catalogs a set of tactics, where each tactic catalogs a set of corresponding techniques, and each technique catalogs a set of corresponding procedure(s). The procedures are collected from publicly available articles and reports describing cyberattacks. ATT\&CK catalogs a set of cyber-criminals or entities performing malicious campaigns tracked by a common name, and we refer to these entities as \textit{cybercrime-groups}~\cite{attack-groups-def}. ATT\&CK also catalogs a set of software/tools, scripts, and executables used for malicious purposes, and we refer to these as \textit{malware}~\cite{attack-malware-def}. ATT\&CK catalogs 125 cybercrime groups and 544 malware which we refer as to \textit{adversary entities}.

\subsection{Mapping between NIST SP800-53 controls and ATT\&CK techniques}
\label{sec:concept-mapping}
Center for Threat Informed Defense (CTID)~\cite{CTID} operated by MITRE Engenuity~\cite{engenuity}, developed a mapping between ATT\&CK version 10 techniques and NIST SP800-53 revision 5 controls. The mapping contains information on whether a given control can mitigate a given technique. For example, \textit{the control named AC-10: Concurrent Session Control mitigates the technique T1185: Browser Session Hijacking}~\cite{T1185}. MITRE ATT\&CK provides a set of mitigation guidelines for techniques. CTID investigates these mitigation guidelines and then identifies the corresponding security controls associated with the guidelines through qualitative analysis and thus performs the mapping. Although the mapping implies whether a particular control can mitigate a particular technique, the mapping cannot indicate: (a) if the control fully or partially mitigates the techniques; (b) if the control is \textit{sufficient} for mitigation of the techniques or the control along with other controls performing synergistically are needed to mitigate the technique. Hence, in this study, if a control \textit{mitigates} a technique, we assume that the control can aid an organization \textit{fully or partially} in mitigating the technique from the aspect of (a) identifying or detecting the techniques; (b) protecting the relevant asset from the impact of the techniques; and (c) responding to or recovering from the damage caused by technique. 

\section{Security Control Metric Suite}
\label{sec:metrics}
Organizations use security controls as a defense mechanism against adversaries. Quantification of the defensive capability of controls can aid in making an informed choice on the controls. Pendleton et al.~\cite{pendletonSurveySystemsSecurity2017} catalog a set of defensive criteria for measuring the strength of defense systems, and we use three following criteria proposed in~\cite{pendletonSurveySystemsSecurity2017}: coverage, redundancy, and risk. The coverage criteria reflect a control's ability to mitigate particular actions of adversaries and a fraction of adversaries. The redundancy criteria reflect the availability of a control alternative, where the alternatives can mitigate the same threat in case of failure. Organizations enforce security controls to mitigate risks from potential cyberthreats. In this study's scope, adversarial techniques pose risks, and security controls reduce or eliminate risk through mitigating techniques. Based upon the three criteria, we propose a security control metric suite containing five metrics to measure coverage, and one metric each to measure redundancy and risk. We discuss the metrics below.

\subsection{Technique Coverage (TEC)} The technique coverage of a control $c$ is the number of techniques $c$ can mitigate based on the mapping~\cref{sec:concept-mapping}. We provide an example. Assume two controls $c_1, c_2$, and three techniques $te_1, te_2, te_3$. $c_1$ mitigates $te_1, te_2$, and $c_2$ mitigates $te_3$. Hence, $TEC(c_1)=2$, and $TEC(c_2)=1$. 

\subsection{Tactic Coverage (TAC)}
\label{def:tac}
Tactic coverage (TAC) of a control $c$ is the percentage of total mitigated techniques of a tactic $ta$. Hence, 

\begin{align}
\label{eqn:tac}
    TAC(c,ta) = \frac{|mte_{ta}|}{|te_{ta}|}
\end{align}

where, $te_{ta}$ is the set of techniques from the tactic $ta$, $|te_{ta}|$ is the number of techniques in $ta$. $mte_{ta} \subseteq te_{ta}$ is the set of mitigated techniques by $c$, and $|mte_{ta}|$ is the number of mitigated techniques from $ta$. We provide an example. Assume two tactics $ta_1,ta_2$, three techniques $te_1, te_2, te_3$ where $te_1, te_3$ are from $ta_1$, and $te_2$ is from $ta_2$. We also assume two controls $c_1, c_2$ where $c_1$ mitigates $te_1, te_2$, and $c_2$ mitigates $te_3$. Hence, 

\begin{align*}
    TAC(c_1, ta_1) = 0.5, TAC(c_1, ta_2) = 1 \\
    TAC(c_2, ta_1) = 0.5, TAC(c_2, ta_2) = 0
\end{align*}

\subsection{Mean Tactic Coverage (MTAC)} Mean tactic coverage (MTAC) of a control $c$ is the mean of TAC of all tactics. Hence, 

\begin{align}
\label{eqn:mtac}
    MTAC(c) = \frac{\sum_{i=1}^{n}TAC(c, ta_i)}{n}
\end{align}

where the number of tactics is $n$. From the example in~\cref{def:tac}:

\begin{align*}
    MTAC(c_1) = \frac{ TAC(c_1, ta_1) + TAC(c_1, ta_2) }{2} = 0.75 \\
    MTAC(c_2) = \frac{ TAC(c_2, ta_1) + TAC(c_2, ta_2) }{2} = 0.25
\end{align*}

\subsection{Control Redundancy (CR)} Control redundancy (CR) of a control $c$ refers to the average number of alternative controls for each of the adversary techniques $c$ mitigates. Hence, 

\begin{align}
\label{eqn:cr}
    CR(c) = \frac{ \sum_{i=0}^{n} ac_i }{n}
\end{align}

where $n$ is the number of adversary techniques $c$ mitigates, $ac_i$ is the count of alternative controls that mitigate the $i^{th}$ technique. We provide an example. We assume, three techniques $te_1, te_2, te_3$ and three controls $c_1, c_2, c_3$. $c_1$ mitigates $te_1, te_2$. $c_2$ mitigates $te_2, te_3$. $c_3$ mitigates $te_3$. We observe that only $c_1$ can mitigate $te_1$, no other control exists that can mitigate $te_1$. However, $te_2$ can be mitigated by both $c_1, c_2$. Hence, $CR(c_1) = \frac{0+1}{2} = 0.5$. $c_2$ mitigates two techniques $te_2, te_3$, and these two techniques have alternative controls for mitigation. Hence, $CR(c_2) = \frac{1+1}{2} = 1$. $c_3$ mitigates one technique $te_3$, and the technique also has one alternative. Hence, $CR(c_3) = \frac{1}{1} = 1$. 

\begin{figure*}[htb]
    \centering
    \includegraphics[width=\linewidth]{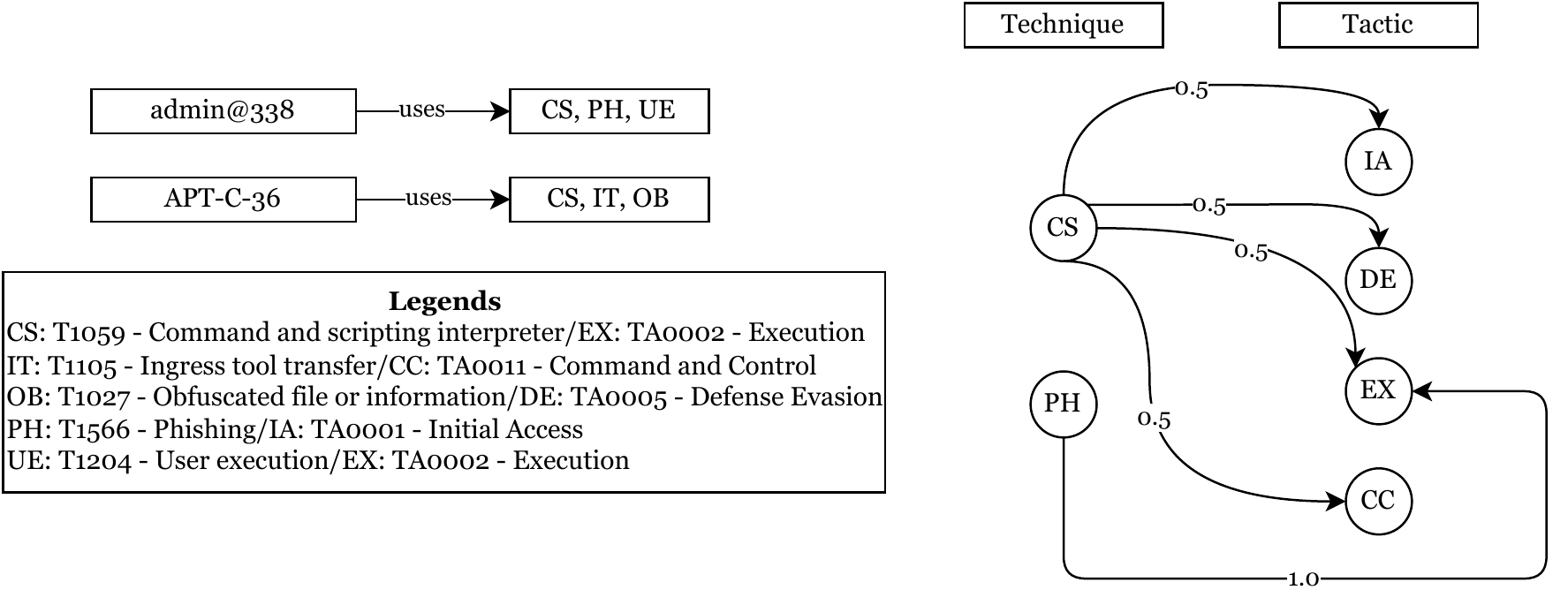}
    \caption{Example of computing risk of adversarial techniques}
    \label{fig:riskExample}
\end{figure*}

\subsection{Adversary Coverage (AC)}
\label{def:ac}
Adversary Coverage (AC) of a control $c$ refers to the fraction of adversary entities where $c$ mitigates at least one technique from the set of techniques used by the adversary entity. Hence, 

\begin{align}
\label{eqn:ac}
    AC(c) = \frac{|mae|}{|ae|}
\end{align}

where $|ae|$ is the count of adversary entities and $|mae|$ is the count of adversaries where at least one technique of the adversary is mitigated by $c$. We provide an example. Assume, four adversary entities $ae_1, ae_2, ae_3, ae_4$ and five techniques $te_1, te_2, te_3, te_4, te_5$. $ae_1$ uses $te_1, te_2$. $ae_2$ uses $te_3, te_4$. $ae_3$ uses $te_5, te_1$. $ae_4$ uses $te_1$. We also assume two controls $c_1, c_2$ where $c_1$ mitigates $te_1, te_2$, and $c_2$ mitigates $te_3, te_4, te_5$. We see, $c_1$ mitigates at least one technique of $ae_1, ae_3, ae_4$, and $c_2$ mitigates at least one technique of $ae_2, ae_3$. Hence, $AC(c_1)=0.75$, and $AC(c_2)=0.5$. 

\subsection{Adversary Technique Coverage (ATC)} Adversary technique coverage (ATC) of a control $c$ refers to the mean count of mitigated techniques of adversary entities. Hence, 

\begin{align}
\label{eqn:atc}
    ATC(c) = \frac{ \sum_{i=1}^{n} pmtae_i }{n}
\end{align}

where $n$ is the number of adversary entities, and $pmtae_i$ denotes the percentage of mitigated techniques by $c$ in $i^{th}$ adversary entity. From the example in~\cref{def:ac}: Adversary Coverage, we observe, $c_1$ mitigates 2 out of 2 techniques of $ae_1$, 0 out of 2 techniques of $ae_2$, 1 out of 2 techniques of $ae_3$, and 1 out of 1 techniques of $ae_4$. $c_2$ mitigates 0 out of 2 techniques of $ae_1$, 2 out of 2 techniques of $ae_2$, 1 out of 2 techniques of $ae_3$, and 0 out of 1 techniques of $ae_4$.  

\begin{align*}
    ATC(c_1) = \frac{ (2/2) + (0/2) + (1/2) + (1/1) }{4} = 62.5 \\
    ATC(c_2) = \frac{ (0/2) + (2/2) + (1/2) + (0/1) }{4} = 37.5
\end{align*}

\subsection{Control's mitigated risk (CMR)}
Control's mitigated risk (CMR) of a control $c$ refers to the mitigated risk of all techniques mitigated by $c$. Thus,

\begin{align}
\label{eqn:cmr}
    CMR(c) = \sum_{i=0}^{n} R(te_i)
\end{align}

where $R(te_i)$ denotes the risk of $i^{th}$ technique among the $n$ mitigated techniques by $c$. We use the following generic equation~\cite[p.~13]{hubbardHowMeasureAnything2016} to measure the risk of an adversarial technique $te$.

\begin{align}
\label{eqn:genericRisk}
    R(te) = LK(te) \times SV(te)
\end{align}

where, $LK(te)$, and $SV(te)$ refers to the likelihood and severity of $te$. We approximate $LK(te)$ by computing the fraction of adversary entities that use the technique. However, we also require to approximate how \textit{severe} an adversarial technique is. We provide an example followed by the equation on how we approximate the severity of adversarial techniques below. 

Fig.~\ref{fig:riskExample} shows a real-world example of adversaries using techniques. A cybercriminal group, admin@338~\cite{admin338}, uses three following techniques: (a) \textit{T1059:}~\cite{T1059}; (b) \textit{T1566}~\cite{T1566}; and (c) \textit{T1204}~\cite{T1204}. Another cybercriminal group named APT-C-36~\cite{aptC36} uses three following techniques: (a) \textit{T1059}; (b) \textit{T1027}~\cite{T1027}; and (c) \textit{T1105}~\cite{T1105}. We observe that adversaries used \textit{T1059} technique with four different adversarial techniques. As adversaries use techniques to achieve tactical goals toward breaching a system, we assume that techniques used in conjunction with techniques from different tactics are relatively more severe. For example, an adversary can perform a plethora of malicious operations through \textit{T1059: Command and scripting interpreter}, such as executing malicious code (\textit{TA0002: Execution}~\cite{TA0002}) or transferring confidential files to remote (\textit{TA0011: Command and control}~\cite{TA0011}). In the figure, adversaries first perform \textit{T1566: phishing} to fulfill \textit{TA0001: Initial access}~\cite{TA0001} and then the adversary tricks users into \textit{T1204: user execution} to gain the tactical objective of \textit{TA0002: Execution}~\cite{TA0002}. Hence, given that adversary uses \textit{T1566: phishing}, the adversary may fulfill the \textit{TA0002: Execution} tactic.

From the example, we observe two severity perspectives of a given technique: (a) how many tactical objectives an adversary gains through using the technique; and (b) how probable an adversary is to gain the tactical objectives in (a) by performing the technique. On the right side of the figure, we build a bipartite graph showing the severity perspective of \textit{T1059} and \textit{T1566}. In the graph, a node on the left represents a technique, a node on the right represents a tactic and an edge represents the technique used with another technique from the corresponding tactic. The degree count of \textit{T1059} and \textit{T1566} are four and one, respectively, indicating the number of tactics the technique is used in conjunction. The labels on the  edges of \textit{T1059} show the likelihood of conjunction given that an adversary uses \textit{T1059}. We use the following equation to measure $SV(te)$ adapted from~\cite{hadarCyberDigitalTwin2020}:

\begin{align}
\label{eqn:severity}
    SV(te) = e^{-h/\alpha} \times C
\end{align}

Here, $C$ denotes the degree count of $te$. $h$ denotes the mean of $1-p_i$ where
$te$ is used in conjunction with techniques from $n$ tactics, and $p_i$ is the likelihood of the conjunction of $i^{th}$ among the $n$ tactics. $\alpha$ is a constant, and we set $\alpha = 8$ as mentioned in~\cite{hadarCyberDigitalTwin2020}. We provide an example of the equation from the figure. The degree count of \textit{T1059} is 4, and techniques from four tactics are used in conjunction with a likelihood of 0.5 each. Hence, $h = \frac{(1-0.5) + (1-0.5) + (1-0.5) + (1-0.5)}{4} = 0.5$ and $SV(T1059) = e^{-0.5/8} \times 4 = 3.76$. The degree count of \textit{T1566} is 1, and techniques from one tactic is used in conjunction with a likelihood of 1.0 each.  Hence, $h = \frac{(1-1) + (1-1)}{2} = 0$ and $SV(T1566) = e^{-0/8} \times 1 = 1$. We provide an example on how we can compute CMR based on the figure. Assume two controls $c_1, c_2$ where $c_1$ mitigates \textit{T1059} and \textit{T1566}, and $c_2$ mitigates \textit{T1566}. Both adversaries use \textit{T1059}, and hence, $LK(T1059) = 1$. One out of the two adversaries uses \textit{T1566}. Hence, $LK(T1566) = 0.5$.

\begin{align*}
    R(T1059) = LK(T1059) \times SV(T1059) = 1 \times 3.76 = 3.76 \\
    R(T1566) = LK(T1566) \times SV(T1566) = 0.5 \times 1 = 0.5 \\
    CMR(c_1) = R(T1059) + R(T1566) = 4.26 \\
    CMR(c_2) = R(T1566) = 0.5
\end{align*}

\textbf{Summary of the metrics:} In Table~\ref{tab:metrics}, we discuss the practical implication of the metrics. 

\begin{table*}[]
    \centering
    \scriptsize
    \caption{Practical implication of the introduced metrics}
    \label{tab:metrics}
    \begin{tabular}{p{3cm}p{14cm}}
    \toprule
    \textbf{Metric} & \textbf{Implication} \\ \midrule
    
        Technique Coverage (TEC) & TEC indicates how many techniques a control can mitigate. A relatively higher TEC of a control indicates that the control mitigates a relatively higher number of techniques. Organizations can achieve protection against a relatively high number of adversarial techniques by implementing controls with high TEC \\ \midrule
        
        Tactic Coverage (TAC) & TAC indicates how many techniques of a given tactic a control can mitigate. A relatively higher TAC of a control and a given tactic indicates that the control mitigates a relatively higher number of techniques of the tactic. An adversary can use multiple techniques of the same tactic to achieve the tactical goal. For example, ATT\&CK catalogs 13 techniques that can aid an adversary in gaining the objective of \textit{TA0004: Privilege escalation}~\cite{ta0004}. Hence, a control having a relatively higher TAC value for \textit{TA0004: Privilege escalation} tactic can aid organizations in gaining relatively higher protection against privilege escalation attempts from adversaries. \\ \midrule
        
        Mean Tactic Coverage (MTAC) & MTAC reflects a control's overall mitigation of techniques from all 14 tactics cataloged in ATT\&CK. A higher value in MTAC indicates the control mitigates a relatively higher number of techniques from all 14 tactics. Hence, organizations can achieve protection against a relatively high number of adversarial techniques across all 14 tactics by implementing controls with high MTAC \\ \midrule
        
        Control Redundancy (CR) & CR reflects how important a control is for mitigating techniques from the perspective of available alternative controls. A relatively lower value indicates that the control has a relatively lower number of alternatives. Organizations may require to enforce the control and ensure that control does not fail \\ \midrule
        
        Adversary Entity Coverage (AC) & AC reflects how relevant a control is against all the cataloged adversary entities in ATT\&CK. A relatively higher value indicates that the control can be effective against a relatively higher number of adversaries. Organizations can achieve protection against a relatively high number of adversaries by implementing controls with high AC  \\ \midrule
        
        Adversary Entity Technique Coverage (ATC) & A control can mitigate multiple techniques, and an adversary can use multiple techniques. Hence, ATC reflects the mean percentage of mitigated techniques used by all cataloged adversaries in ATT\&CK. Organizations can achieve protection against a relatively high percentage of techniques of all adversaries by implementing controls with high ATC  \\ \midrule
        
        Control's Mitigated Risk (CMR) & CMR reflects the risk mitigation capability of a control. Organizations can mitigate a relatively higher amount of risk by implementing controls with a relatively higher CMR \\ 
    \bottomrule
    \end{tabular}
\end{table*}

\section{Methodology}
\label{method}
In this section, we discuss the methodology of the study. 

\begin{figure}
    \centering
    \includegraphics[width=\linewidth]{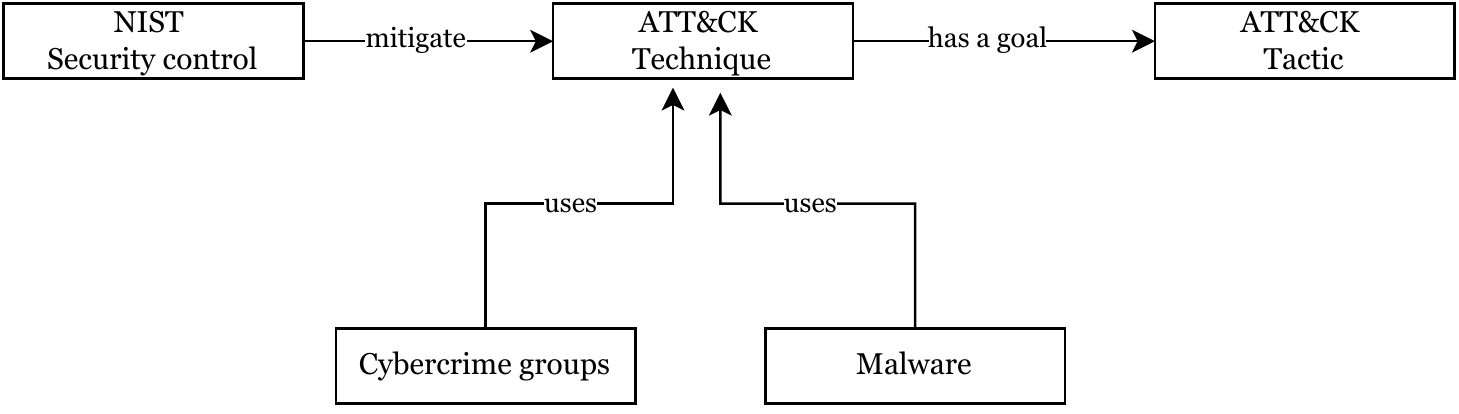}
    \caption{Dataset Schema}
    \label{fig:schema}
\end{figure}

\textbf{Construct dataset:} We obtain the catalog of NIST SP800-53r5 controls from~\cite{NIST800r53}. We then obtain the following from ATT\&CK webpage~\cite{working-with-attack}: (a) catalog of ATT\&CK tactics from~\cite{dbattacktactics}; (b) catalog of ATT\&CK techniques from~\cite{db-attack-techniques}; (c) catalog of cybercrime groups and the set of techniques used by each group from~\cite{db-attack-groups}; and (d) catalog of malware and the set of techniques used by each malware from~\cite{db-attack-malware}. The obtained MITRE ATT\&CK catalogs are from Enterprise MITRE ATT\&CK version 10. We then obtain the mapping between NIST SP800-53 controls and ATT\&CK techniques from~\cite{attack-control-mapping}. Based on the mapping, we combine the obtained data according to the schema presented in Fig.~\ref{fig:schema}. According to the schema: (a) adversary entities use ATT\&CK techniques; (b) adversaries achieve ATT\&CK tactic by using ATT\&CK techniques; (c) NIST SP800-53 controls mitigate ATT\&CK techniques. The dataset contains information on 298 controls, 188 ATT\&CK techniques, 14 ATT\&CK tactics. The dataset also contains the set of techniques used by 669 adversary entities, among which 125 entities are cybercrime groups, and 544 entities are malware. 

\textbf{Evaluate controls on MITRE ATT\&CK tactics, techniques, and adversary entities:} We first evaluate the controls based on the TEC, TAC, MTAC, and CR metrics from the proposed metric suite. We perform the evaluation on cataloged adversarial techniques in MITRE ATT\&CK based on the mapping~\cref{sec:concept-mapping}. We then evaluate the controls based on AC, ATC, and CMR metrics on the cataloged adversary entities in MITRE ATT\&CK based on the mapping~\cref{sec:concept-mapping}. We answer RQ1 based on the evaluation of controls by the aforementioned metrics.

\textbf{Apply clustering:} Based upon the evaluation of NIST SP800-53 controls on ATT\&CK tactics, techniques, and adversary entities, we obtain numeric values of TEC, MTAC, AC, ATC, CR, and CMR of all the controls that can mitigate techniques. First, we normalize all the values for the corresponding metrics from 0 to 1. As six dimensions exist for each control, we apply principal component analysis~\cite{abdi2010principal} to reduce the dimensions of the controls into 2. We then perform clustering on the controls. The clustering groups the controls having similar characteristics into clusters based on the metrics, and we then answer RQ2 by observing the characteristics of each of the obtained clusters. We first apply the elbow method~\cite{thorndike1953belongs} to identify the optimal number of clusters, and we then apply K-Means clustering~\cite{Lloyd1982Least} to identify the clusters among the controls. 

\section{Findings on RQ1}
\label{sec:rq1}
This section reports our findings on evaluating NIST SP800-53 controls by TEC, TAC, MTAC, and CR metrics on the mapped MITRE ATT\&CK techniques and corresponding tactics. Then we report our findings on evaluating NIST SP800-53 controls by AC, ATC, and CMR metrics on 669 adversary entities cataloged in MITRE ATT\&CK. 

\begin{table}[]
    \centering
    \scriptsize
    \caption{Top ten controls by TEC metric}
    \label{tab:control-tec}
    \begin{tabular}{lr}
    \toprule
        \textbf{Security Control} & \textbf{TEC} \\ \midrule
        SI-4: System Monitoring & 120 \\
        CM-6: Configuration Settings & 111 \\
        CM-2: Baseline Configuration & 97 \\
        AC-3: Access Enforcement & 91 \\
        CM-7: Least Functionality & 85 \\
        CA-7: Continuous Monitoring & 84 \\
        AC-6: Least Privilege & 83 \\
        SI-3: Malicious Code Protection & 81 \\
        SI-7: Software, Firmware, and Information Integrity & 70 \\
        SC-7: Boundary Protection & 68 \\ \midrule 
        Number of controls with $TEC > 0$ & 101 \\
        Q1 and Q3 ($25^{th}$ and $75^{th}$ percentile) of TEC of mitigating controls & 3, 19 \\
        Mean and median of TEC of mitigating controls & 19, 6 \\
    \bottomrule
    \end{tabular}
\end{table}

\subsection{Findings on Technique Coverage (TEC)}
\label{sec:rq1-tec}
We report the top ten controls by TEC values in Table~\ref{tab:control-tec}. We observe that the top ten controls mitigate between 36\% to 64\% of the 188 techniques in ATT\&CK. From the top ten, we observe the following aspects of security functions that can aid organizations in gaining protection from a relatively high number of adversary techniques. First, organizations need to implement system-wide monitoring of operations across networks, devices, applications, data, and users (SI-4). Second, organizations need to establish and document secure configurations across system components that they can review, test, and improve in a traceable manner (CM-6). Finally, organizations need to enforce the least-required functional privilege across users, applications, and devices with appropriate authentication and authorization (AC-6, CM-7). From the table, we also observe that only 101 (34\%) of 298 NIST controls have a TEC value greater than zero which indicates that the 101 controls can mitigate at least one technique. In contrast, the remaining 197 controls cannot mitigate any technique. We refer to these 101 controls as \textit{mitigating controls}. The mean, median, Q3 of the TEC values suggest that the mitigating controls primarily mitigate a relatively lower number of techniques. However, the top ten mitigates techniques three to six times more than that of controls in the $75^{th}$ percentile.  The observation also emphasizes that organizations should prioritize the enforcement of the top ten controls shown in the table because of their ability to mitigate a relatively large number of techniques.

\begin{table}[]
    \centering
    \scriptsize
    \caption{Top ten techniques having the most number of controls}
    \label{tab:technique-tec}
    \begin{tabular}{lr}
    \toprule
        \textbf{Technique} & \textbf{Count} \\ \midrule
        T1552: Unsecured Credentials - TA0006: Credential access & 35 \\
        T1530: Data from Cloud Storage Object - TA0009: Collection & 33 \\
        T1213: Data from Information Repositories - TA0009: Collection & 31 \\
        T1021: Remote Services - TA0008: Lateral Movement & 31 \\
        T1210: Exploitation of Remote Services - TA0008: Lateral Movement & 30 \\
        T1547: Boot or Logon Autostart Execution - TA0003: Persistence & 29 \\
        T1190: Exploit Public-Facing Application - TA0001: Initial Access & 27 \\
        T1574: Hijack Execution Flow - TA0005: Defense Evasion & 27 \\
        T1602: Data from Configuration Repository - TA0009: Collection & 25 \\
        T1505: Server Software Component - TA0003: Persistence & 25 \\ \midrule
        Number of mitigated techniques & 135 \\
        Q1 and Q3 of the count of controls of mitigated techniques & 8, 20 \\
        Mean and median of the count of controls of mitigated techniques & 14, 13 \\
    \bottomrule
    \end{tabular}
\end{table}

We report the top ten techniques having the highest number of controls for mitigation in Table~\ref{tab:technique-tec} and descriptive statistics on the count of the mitigating controls of the techniques. We show the names of the techniques followed by the corresponding tactic. The table shows that 135 out of 188 ATT\&CK techniques have at least one mitigating control, indicating that 53 (28\%) of ATT\&CK techniques cannot be mitigated by any control. We refer to the 135 techniques having at least one mitigating control as the \textit{mitigable techniques}. The mean and median of the mitigable techniques indicate that the mitigable techniques primarily have multiple mitigating controls. As we mention in the~\cref{sec:concept-mapping}, despite a technique having multiple mitigating controls, the technique may not be mitigated by only a single control. Hence, organizations may need to implement more than one control to mitigate any techniques. On the other hand, a technique with multiple mitigating controls suggests that the mitigation of the technique may require the combined enforcement of multiple controls. For example, the topmost technique in Table~\ref{tab:technique-tec}, \textit{T1552: Unsecured Credentials}, has 35 mitigating controls. Except \textit{SI-3}, all the top ten controls in Table~\ref{tab:control-tec} mitigate this technique. The observation suggests that enforcing access control or applying system monitoring cannot single-handedly deter an adversary from exploiting unsecured credentials. We observe that, except for \textit{T1547}, and \textit{T1574}, an adversary can exploit the eight other techniques remotely to collect confidential data, perform a lateral movement, or gain initial access. All the top ten techniques shown in Table~\ref{tab:technique-tec} have at least 25 mitigating controls. The observation emphasizes that organizations need to enforce controls to protect the remote facets of their systems most.


\begin{table}[]
    \centering
    \scriptsize
    \caption{Mitigating controls in each tactic category}
    \label{tab:tactic-control-count}
    \begin{tabular}{lrR{15mm}r}
    \toprule
        \textbf{Tactic name} & \textbf{CT} & \textbf{CPMT} & \textbf{CC} \\ \midrule
        TA0001: Initial Access~\cite{TA0001} & 9 & 9 (100\%) & 54 \\
        TA0002: Execution~\cite{TA0002} & 12 & 12 (100\%) & 47 \\
        TA0003: Persistence~\cite{TA0003} & 19 & 19 (100\%) & 58 \\
        TA0004: Privilege escalation~\cite{ta0004} & 13 & 13 (100\%) & 59 \\
        TA0005: Defense evasion~\cite{TA0005} & 40 & 31 (78\%) & 69 \\
        TA0006: Credential access~\cite{TA0006} & 15 & 15 (100\%) & 54 \\
        TA0007: Discovery~\cite{TA0007} & 29 & 10 (34\%) & 29 \\
        TA0008: Lateral movement~\cite{TA0008} & 9 & 8 (89\%) & 54 \\
        TA0009: Collection~\cite{TA0009} & 17 & 11 (65\%) & 57 \\
        TA0010: Exfiltration~\cite{TA0010} & 9 & 9 (100\%) & 39 \\
        TA0011: Command and control~\cite{TA0011} & 16 & 16 (100\%) & 26 \\
        TA0040: Impact~\cite{TA0040} & 13 & 10 (77\%) & 39 \\ 
        TA0042: Resource Development~\cite{TA0042} & 7 & 0 (0\%) & 0 \\
        TA0043: Reconnaissance~\cite{TA0043} & 10 & 1 (10\%) & 11 \\ \midrule
        \multicolumn{4}{p{8cm}}{Legends: CT: count of techniques, CPMT: count and percentage of mitigated techniques, CC: count of mitigating controls  } \\
        \bottomrule
    \end{tabular}
\end{table}

\setlength{\tabcolsep}{0.5em}
\begin{table*}[htb]
    \centering
    \scriptsize
    \caption{Top ten controls by MTAC and TAC of 14 ATT\&CK tactics}
    \label{tab:mtac}
    
    \begin{tabular}{p{40mm}rrrrrrrrrrrrrrr}
    \toprule
    {} & \multicolumn{14}{c}{\textbf{TAC}} & {}  \\ \cmidrule{2-15}
     \textbf{Control}                                             &   \textbf{IA} &   \textbf{EX} &   \textbf{PS} &   \textbf{PE} &   \textbf{DE} &   \textbf{CA} &   \textbf{DC} &   \textbf{LM} &   \textbf{CL} &   \textbf{EF} &   \textbf{CC} &   \textbf{IM} &   \textbf{RD} &   \textbf{RC} &   \textbf{MTAC} \\
    \midrule
     SI-4: System Monitoring                             &     0.67 &     *0.92 &     0.95 &     0.92 &     *0.7  &     *0.93 &     *0.21 &     *0.89 &     *0.65 &     *1    &     *1    &     0.62 &        0 &      *0.1 &   0.68 \\
     CM-6: Configuration Settings                        &     *0.78 &     0.83 &     *1    &    *1    &     *0.7  &     0.87 &     *0.21 &     0.78 &     0.41 &     0.78 &     0.94 &     0.46 &        0 &      *0.1 &   0.63 \\
     AC-3: Access Enforcement                            &     0.67 &     0.67 &     0.84 &     0.92 &     0.57 &     0.73 &     0.17 &     *0.89 &     0.53 &     0.56 &     0.38 &     *0.77 &        0 &      0   &   0.55 \\
     CM-2: Baseline Configuration                        &     0.44 &     0.75 &     0.84 &     0.77 &     0.55 &     0.8  &     0.07 &     *0.89 &     0.41 &     0.67 &     0.94 &     0.46 &        0 &      *0.1 &   0.55 \\
     CA-7: Continuous Monitoring                         &     0.56 &     0.5  &     0.63 &     0.77 &     0.48 &     0.87 &     0.07 &     0.78 &     0.35 &     0.78 &     0.94 &     0.31 &        0 &      *0.1 &   0.51 \\
     AC-6: Least Privilege                               &     *0.78 &     0.83 &     0.84 &     0.92 &     0.57 &     0.67 &     0.17 &     0.67 &     0.35 &     0.56 &     0    &     0.54 &        0 &      0   &   0.49 \\
     CM-7: Least Functionality                           &     0.44 &     *0.92 &     0.84 &     0.77 &     0.52 &     0.4  &     0.17 &     0.67 &     0.24 &     0.44 &     0.81 &     0.38 &        0 &      0   &   0.47 \\
     SI-3: Malicious Code Protection                     &     0.44 &     0.67 &     0.53 &     0.69 &     0.3  &     0.47 &     0.07 &     0.67 &     0.41 &     0.78 &     0.94 &     0.38 &        0 &      *0.1 &   0.46 \\
     AC-4: Information Flow Enforcement                  &     0.56 &     0.33 &     0.42 &     0.46 &     0.25 &     0.4  &     0.07 &     0.56 &     0.29 &     0.78 &     0.94 &     0.31 &        0 &      *0.1 &   0.39 \\
     SI-7: Software, Firmware, and Information Integrity &     0.44 &     0.83 &     0.79 &     0.69 &     0.5  &     0.53 &     0.03 &     0.44 &     0.47 &     0.11 &     0.06 &     0.54 &        0 &      0   &   0.39 \\ \midrule
     
     Mean & 0.15 & 0.15 & 0.18 & 0.20 & 0.12 & 0.16 & 0.02 & 0.18 & 0.12 & 0.14 & 0.10 & 0.09 & 0.00 & 0.01 & 0.11 \\
     
     Median & 0.11 & 0.00 & 0.05 & 0.08 & 0.05 & 0.07 & 0.00 & 0.11 & 0.06 & 0.00 & 0.00 & 0.00 & 0.00 & 0.00 & 0.04 \\
     
    \midrule
    \multicolumn{16}{p{17cm}}{IA: TA0001 - Initial access, EX: TA0002 - Execution, PS: TA0003 - Persistence, PE: TA0004 - Privilege escalation, DE: TA0005 - Defense evasion, CA: TA0006 - Credential access, DC: TA0007 - Discovery, LM: TA0008 - Lateral movement, CL: TA0009 - Collection, EF: TA0010 - Exfiltration, CC: TA0011 - Command and control, IM: TA0040 - Impact, RD: TA0042 - Resource development, RC: TA0043 - Reconnaissance  } \\
    \bottomrule
    \end{tabular}
    
\end{table*}

\subsection{Findings on Tactic Coverage (TAC) and Mean Tactic Coverage (MTAC)}
We first report the number of mitigated techniques and mitigating controls in each of the 14 tactic categories in Table~\ref{tab:tactic-control-count}. We observe that techniques belonging to \textit{Resource development} tactic do not have any mitigating controls. We also observe that seven tactics have all their respective techniques mitigated. However, we identify six tactics where not all of the techniques are mitigated. For example, the \textit{Reconnaissance} tactic has ten techniques but only one of them can be mitigated. Similarly, 66\% of the techniques in the \textit{Discovery} tactic are not mitigated. The number of mitigating controls are highest in the three following tactics: \textit{Defense evasion}, \textit{Persistence}, and \textit{Privilege escalation}. However, the following tactics have the lowest number of mitigating controls: \textit{Reconnaissance}, \textit{Command and control}, and \textit{Discovery}, although \textit{Command and control} have all their techniques mitigated. The observations suggest that new controls should be identified for \textit{Discovery}, \textit{Defense evasion}, \textit{Collection}, and \textit{Impact} tactics as these tactics have unmitigated techniques. On the other hand, only one technique: \textit{T1598: Phishing for information} is mitigable out of 17 techniques from the \textit{Reconnaissance} and \textit{Resource development} tactics.  However, adversaries use techniques from these two tactics to formulate cyberattack strategies, acquire, and configure resources on their end. Thus, adversaries do not directly operate on the victim's end. Hence, developing controls to mitigate these techniques from the two tactics is practically infeasible. Organizations should limit the availability of the organizations and systems-related sensitive information to third parties, as suggested in ATT\&CK~\cite{mitigationM1056}.

We report the top ten controls by MTAC metric and corresponding TAC metric for the 14 ATT\&CK tactics in Table~\ref{tab:mtac}. We indicate the highest TAC value among all mitigating controls for a given tactic by an asterisk. In case of a tie, we give an asterisk to all the highest values. First, we observe that, except \textit{AC-4}, all other controls also appear in the top ten control by TEC metric (Table~\ref{tab:control-tec}). The observation suggests that the more techniques a control mitigates, the better coverage the control has across all 14 tactics. We also observe that \textit{SI-4} and \textit{CM-6} mitigate the highest percentage of techniques of 9 and 6 tactics, respectively. \textit{AC-3} and \textit{CM-2} have the highest TAC value for two tactics, while \textit{AC-6, CM-7} and \textit{CA-7} have the highest TAC value for only one tactic each. The observation suggests that system-wide monitoring, secure configuration, and appropriate access with the least privilege/functionality of system components can be the most effective enforcement against multiple adversary tactics. The highest mean and median of the fourteen TAC values across all mitigating controls are 0.20 and 0.11, respectively. The highest mean and median MTAC values across all 101 mitigating controls are 0.11 and 0.04, respectively. The observations indicate the top ten controls in Table~\ref{tab:mtac} mitigate a relatively high number of techniques from 14 tactics. However, a relatively low mean and lower median TAC for 14 tactics and MTAC suggest that a control primarily does not mitigate a high percentage of techniques from a single tactic and does not have a high MTAC value. Hence, organizations may enforce the top ten controls in Table~\ref{tab:mtac} as a priority to achieve overall protection against various adversary tactics. 

\begin{table}[htb]
    \centering
    \scriptsize
    \caption{Top ten controls having the least CR value }
    \label{tab:cr}
    \begin{tabular}{p{6cm}rr}
    \toprule
        \textbf{Control} & \textbf{CR} & \textbf{TEC} \\
        
        AC-8: System Use Notification & 6.00 & 1 \\
        MA-5: Maintenance Personnel & 6.00 & 1 \\
        CP-2: Contingency Plan & 9.60 & 5 \\
        MP-7: Media Use & 10.00 & 5 \\
        SC-41: Port and I/O Device Access & 10.75 & 4 \\ 
        
        CP-10: System Recovery and Reconstitution & 11.83 & 6 \\
        SC-21: Secure Name/address Resolution Service & 12.00 & 2 \\
        SC-22: Architecture and Provisioning for Name/address Resolution Service & 12.00 & 2 \\
        SC-38: Operations Security & 12.00 & 2 \\
        SI-8: Spam Protection & 12.20 & 5 \\ \midrule
        
        \multicolumn{3}{l}{Mean and median CR of mitigating controls: 18.2, and 18.7} \\ 
        
    \bottomrule
    \end{tabular}
\end{table}

\begin{figure}
    \centering
    \includegraphics[width=\columnwidth]{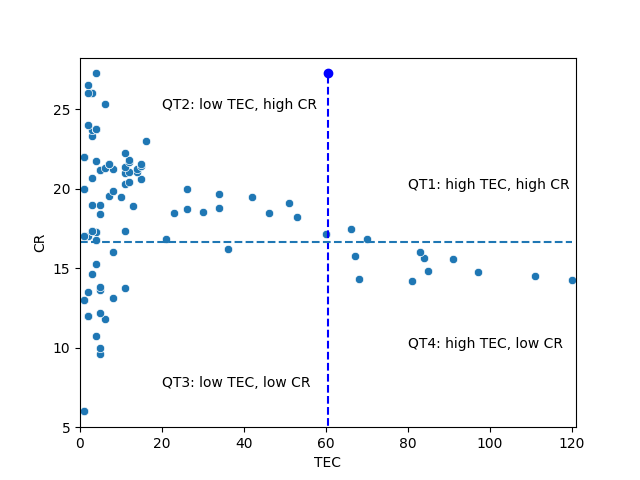}
    \caption{Scatter plot of TEC and CR of mitigating controls}
    \label{fig:plot-cr-tec}
\end{figure}

\subsection{Findings on Control Redundancy (CR)}
We report the top ten controls having the least CR values in Table~\ref{tab:cr}. Although a relatively low CR value indicates that the corresponding controls have a relatively lower number of alternatives, we observe in Table~\ref{tab:cr} that the top ten controls mitigate a relatively lower number of techniques as well. Five out of the ten controls mitigate two techniques at most while the remaining ones mitigate at most six techniques. We provide a scatter plot of TEC and CR of controls in Fig.~\ref{fig:plot-cr-tec} to gain further perspective on the TEC and CR of the controls. We divide the plot into four equal quarters. We observe ten controls with relatively high TEC but relatively low CR values (QT4). Organizations can prioritize the enforcement of these controls as they have relatively fewer alternative controls but can mitigate a relatively higher number of techniques. These ten controls are the top nine controls shown in Table~\ref{tab:mtac}, and SC-7: Boundary Protection. On the other hand, we identify 67 controls with relatively high CR but relatively low TEC values (QT2). Hence, these controls can have lower priority as they have a relatively higher number of alternatives but do not mitigate many techniques. Finally, we identify 23 controls with either high TEC and CR values (QT1) or low TEC and CR values (QT3). Organizations can also prioritize these controls as they either mitigate a large number of techniques or have a lower number of alternative controls.

\begin{table}[htb]
    \centering
    \scriptsize
    \caption{Top Ten controls by AC and ATC}
    \label{tab:ac-atc}
    \begin{tabular}{lrr}
    \toprule
    \textbf{Control} & \textbf{AC} & \textbf{ATC} \\ \midrule
        SI-4: System Monitoring & 0.98 & 0.71(1) \\
        CM-6: Configuration Settings & 0.98 & 0.66(2) \\
        CM-2: Baseline Configuration & 0.96 & 0.61(4) \\
        SI-3: Malicious Code Protection & 0.96 & 0.62(3) \\
        CA-7: Continuous Monitoring & 0.96 & 0.57(5) \\
        CM-7: Least Functionality & 0.95 & 0.51(6) \\
        AC-3: Access Enforcement & 0.94 & 0.47(7) \\
        AC-6: Least Privilege & 0.92 & 0.42(8) \\
        SI-7: Software, Firmware, and Information Integrity & 0.92 & 0.41(9) \\
        AC-2: Account Management & 0.90 & 0.37(10) \\ \midrule
        
        \multicolumn{3}{l}{Mean and median of AC of mitigating controls: 0.42, 0.38 } \\
        \multicolumn{3}{l}{Mean and median of ATC of mitigating controls: 0.11, 0.04 } \\ 
        
    \bottomrule
    \end{tabular}
\end{table}

\subsection{Findings on Adversarial Coverage (AC) and Adversarial Technique Coverage (ATC)}
We report the top ten controls by the AT and ATC metrics in Table~\ref{tab:ac-atc}. The controls are ranked by the AC metric. For the ATC metric, we show the rank after the corresponding ATC value. We observe, for both AC and ATC, the top ten controls are the same, and the order is similar except \textit{CM-6, CM-2}, and \textit{SI-3}. We observe, the top nine controls in Table~\ref{tab:ac-atc} are identical to the top nine controls by TEC in Table~\ref{tab:control-tec}. The observation suggests the top ten controls have relatively higher technique coverage; hence, they can solve at least one technique of 90\% of the adversaries. We also observe that the top ten controls can mitigate, on average, 37-71\% of the techniques used by adversaries. However, the mean and median ATC of all mitigating controls is relatively very low compared to that of the ATC of the top ten, which indicates that organizations may prioritize the controls in Table~\ref{tab:ac-atc} to protect against a relatively high percentage of adversaries and to nullify a relatively high percentage of adversarial techniques from each adversary. 

\begin{table}[]
    \centering
    \scriptsize
    \caption{Top Ten most used unmitigated techniques by the adversaries}
    \label{tab:unmitigated-techniques}
    \begin{tabular}{L{75mm}r}
    \toprule
        \textbf{Technique} & \textbf{Pct} \\ \midrule
        T1082:System Information Discovery~\cite{T1082} - TA0007: Discovery & 0.42 \\
        T1083:File and Directory Discovery~\cite{T1083} - TA0007: Discovery & 0.36 \\
        T1057:Process Discovery~\cite{T1057} - TA0007: Discovery & 0.31 \\
        T1016:System Network Configuration Discovery~\cite{T1016} - TA0007: Discovery & 0.29 \\
        T1140:Deobfuscate/Decode Files or Information~\cite{T1140} - TA0005: Defense Evasion & 0.26 \\
        T1033:System Owner/User Discovery~\cite{T1033} - TA0007: Discovery & 0.22 \\
        T1113:Screen Capture~\cite{T1113} - TA0009: Collection & 0.18 \\
        T1518:Software Discovery~\cite{T1518} - TA0007: Discovery & 0.15 \\
        T1074:Data Staged~\cite{T1074} - TA0009: Collection & 0.12 \\
        T1012:Query Registry~\cite{T1012} - TA0007: Discovery & 0.12 \\ \midrule
        
        \multicolumn{2}{p{8cm}}{Mean and median percentage of unmitigated techniques among adversaries: 0.27, 0.26} \\
    \bottomrule
    \end{tabular}
\end{table}

However, as we identify in~\cref{sec:rq1-tec} that ATT\&CK contains 53 techniques that do not have mitigating controls, we investigate to what extent adversaries use the unmitigated techniques. We report the top ten most used unmitigated techniques by the adversaries in Table~\ref{tab:unmitigated-techniques}. The table shows the techniques, their corresponding tactic, and the percentage of adversaries using the technique. We identify that unmitigated techniques are primarily from \textit{TA0007: Discovery} tactic, where adversaries use these techniques to identify software, hardware, networks, applications, operating systems, vulnerabilities, and architecture-related information of the target systems. The information aids adversaries in formulating their attack strategies. The table also suggests that, on average, 25\% of adversarial techniques from an adversary do not have any controls for mitigation. Hence, organizations must introduce new controls for mitigating these techniques and prioritize detection of these techniques in their environments. 

\begin{table}[]
    \centering
    \scriptsize
    \caption{Top Ten techniques having the most risk}
    \label{tab:technique-risk}
    \begin{tabular}{p{65mm}rr}
    \toprule
    \textbf{Techniques} & \textbf{Risk} & \textbf{CC} \\ \midrule
    
    T1059: Command and Scripting Interpreter~\cite{T1059} - TA0002: Execution & 0.52 & 24 \\
    T1105: Ingress Tool Transfer~\cite{T1105} - TA0011: Command and Control & 0.47 & 8 \\
    T1027: Obfuscated Files or Information~\cite{T1027} - TA0005: Defense Evasion & 0.44 & 6 \\
    T1071: Application Layer Protocol~\cite{T1071} - TA0011: Command and Control & 0.40 & 18 \\
    T1082: System Information Discovery~\cite{T1082} - TA0007: Discovery & 0.40 & 0 \\
    T1083: File and Directory Discovery~\cite{T1083} - TA0007: Discovery & 0.34 & 0 \\
    T1070: Indicator Removal on Host~\cite{T1070} - TA0005: Defense Evasion & 0.33 & 20 \\
    T1547: Boot or Logon Autostart Execution~\cite{T1547} - TA0003: Persistence & 0.30 & 29 \\
    T1057: Process Discovery~\cite{T1057} - TA0007: Discovery & 0.29 & 0 \\
    T1016: System Network Configuration Discovery~\cite{T1016} - TA0007: Discovery & 0.27 & 0 \\ \midrule
    
    \multicolumn{3}{l}{Mean and median of risks of techniques: 0.62, 0.02 } \\ 
    \multicolumn{3}{l}{Q1 and Q3 of risks of techniques: 0.01, 0.08} \\ \midrule
    
    \multicolumn{3}{l}{CC: Count of mitigating controls of techniques} \\ 
    
    \bottomrule
    \end{tabular}

\end{table}

\subsection{Findings on Control's Mitigated Risk (CMR)}
We report the top ten techniques having the most risk in Table~\ref{tab:technique-risk}. We observe that the top four techniques are built-in system functionalities abused by the adversaries. For example, through \textit{T1059: Command and scripting interpreter} technique, adversaries use the command line interface, which comes pre-built in all major operating systems. We also observe four techniques from the \textit{TA0007: Discovery} tactic are present in the top ten techniques. However, none of the four has any mitigating control. We report the top ten controls that mitigate the most risks in Table~\ref{tab:control-risk}. We observe that the top nine controls in Table~\ref{tab:control-risk} also appear among the top ten controls shown in Table~\ref{tab:control-tec}. The observation suggests that the controls mitigating a relatively higher number of techniques have a relatively higher value in CMR. The mean and median of CMR of mitigating controls also suggest that a control primarily does not have a relatively high CMR value. Hence, the top controls shown in Table~\ref{tab:control-risk} are more important than the rest in terms mitigating risks incurred by the adversarial techniques. We investigate common controls that mitigate all the six mitigable techniques shown in Table~\ref{tab:technique-risk}, and we identify the four following controls: SI-4: System Monitoring, SI-3: Malicious Code Protection, CM-6: Configuration Settings, and CM-2: Baseline Configuration. Hence, organizations can enforce these four controls with maximum prioritization.  

\begin{table}[]
    \centering
    \scriptsize
    \caption{Top Ten controls by CMR}
    \label{tab:control-risk}
    \begin{tabular}{lrr}
    \toprule
    \textbf{Control} & \textbf{CMR} & \textbf{TEC} \\ \midrule
    
    SI-4: System Monitoring & 8.03 & 120.00 \\
    CM-6: Configuration Settings & 7.37 & 111.00 \\
    SI-3: Malicious Code Protection & 6.96 & 81.00 \\
    CM-2: Baseline Configuration & 6.70 & 97.00 \\
    CA-7: Continuous Monitoring & 6.32 & 84.00 \\
    CM-7: Least Functionality & 5.57 & 85.00 \\
    AC-3: Access Enforcement & 5.31 & 91.00 \\
    AC-6: Least Privilege & 4.88 & 83.00 \\
    SI-7: Software, Firmware, and Information Integrity & 4.51 & 70.00 \\
    AC-2: Account Management & 4.26 & 66.00 \\ \midrule
    
    \multicolumn{3}{l}{Mean and Median CMR of mitigating controls: 1.26, 0.44} \\ 
    \multicolumn{3}{l}{Q1 and Q3 CMR of mitigating controls: 0.18, 1.27} \\ 
    
    \bottomrule
    \end{tabular}

\end{table}

\begin{figure}
    \centering
    \includegraphics[width=\columnwidth]{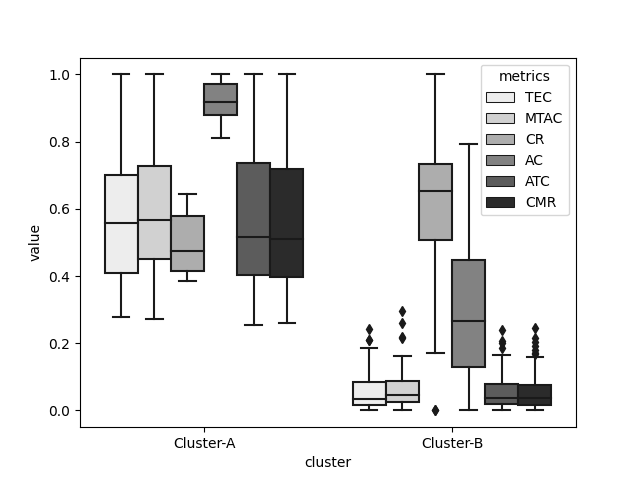}
    \caption{Boxplot of the metrics of the two identified clusters of the mitigating control}
    \label{fig:cluster}
\end{figure}

\section{Findings on RQ2}
\label{sec:rq2}
We perform clustering on the controls based on the six metrics: TEC, MTAC, CR, AC, ATC, and CMR. After applying the elbow method, we identify that the optimal number of clusters is two. Then we apply K-means clustering to identify the two clusters. We provide the boxplot of the six metrics of the two clusters in Fig.~\ref{fig:cluster}. We name the two clusters as \textit{Cluster-A} and \textit{Cluster-B}. \textit{Cluster-A} contains 20 controls in total, while the remaining 81 controls belong to \textit{Cluster-B}. Comparing the two clusters, we observe that controls in \textit{Cluster-A} mitigate a higher number of techniques (TEC), mitigate a higher percentage of techniques across all 14 ATT\&CK tactics (MTAC), and are effective against a higher number of adversary entities(AC and ATC). However, we also identify that controls in the \textit{Cluster-A} have a lower number of redundant controls(CR). Controls in \textit{Cluster-A} also mitigate a higher amount of risk than that of \textit{Cluster-B}. In Table~\ref{tab:cluster-comparison}, we report the comparison between the two clusters and we observe the following. Controls from \textit{Cluster-A} mitigate 23 more techniques than that of \textit{Cluster-B}. Controls from \textit{Cluster-A} mitigate, on average 10\% more techniques across all 14 ATT\&CK tactics than that of \textit{Cluster-B}. Controls from \textit{Cluster-A} mitigate at least one technique in 2\% more adversary entities than that of \textit{Cluster-B}. Finally, controls from \textit{Cluster-A} mitigate, on average, 18\% more techniques from each adversary entity than that of \textit{Cluster-B}. Overall, we observe that, controls from \textit{Cluster-A} outperforms controls from \textit{Cluster-B} despite the fact that, \textit{Cluster-B} contains four times more control than that of \textit{Cluster-A}. Overall, organizations can prioritize the enforcement of the 20 controls in the \textit{Cluster-A} over the remaining 81 controls from the \textit{Cluster-B}. We refer to these controls as the top 20 critical controls and report them in Table~\ref{tab:top20}. From the table, we observe that organizations should focus on the following measures to ensure protection against adversarial techniques. 

\begin{table}[]
    \centering
    \scriptsize
    \caption{Comparison between the two clusters of controls}
    \label{tab:cluster-comparison}
    \begin{tabular}{lrrrrr}
    \toprule
    
    \textbf{Cluster} & \textbf{Controls} & \textbf{MT} & \textbf{AFTM} & \textbf{FA} & \textbf{AFTMA} \\ \midrule
    
    Cluster-A & 20 & 134 & 0.75 & 0.98 & 0.72 \\
    Cluster-B & 81 & 111 & 0.65 & 0.96 & 0.54 \\ \midrule
    
    \multicolumn{6}{p{8cm}}{MT: count of total mitigated techniques, AFTM: average fraction of techniques mitigated from all tactics, FA: fraction of adversaries where at least one technique is mitigated, AFTMA: average fraction of techniques mitigated for all adversaries } \\ 
    
    \bottomrule
    \end{tabular}

\end{table}

\textbf{Secure the access to the system:} Organizations should enforce appropriate security to all possible access paths to the systems (AC-2, AC-3). Organizations should also ensure that all the entities gaining physical or logical access to the systems are authenticated and authorized (IA-2). During the process of introducing any change in system components or its actors, the system may momentarily become vulnerable, and hence, organizations should restrict access to the systems (CM-5). Moreover, the flow of information inside the systems and between external systems should also be secured through appropriate access control, so that no system component or users gains access to the information not required for the roles and responsibilities (AC-4).

\textbf{Apply right configurations:} Information systems primarily consist of heterogeneous components and users which work together to support business activities. The users and components can have different configurations, in the context of what extent of functionalities, privileges, or responsibilities are assigned to them. Oftentimes, a misconfiguration in components and users introduces security weaknesses to a system. For example, an application only requires a \textit{write} permission to perform a task, however, if the application is misconfigured to have \textit{execute} permission, then adversaries can execute malicious code by compromising the application. Thus, organizations should focus on the right configuration of the components they are using to minimize the attack surface. Hence, organizations should take a minimalist approach while assigning duties, privileges, and functionalities to the system components(AC-5, AC-6, CM-7). Organizations should also adopt a baseline security configuration that they can review, test, and update in case of identified compromises (CM-2, CM-6). Moreover, they should maintain a traceable inventory of system components and implement a mechanism for detecting unauthorized changes made to the system (CM-8, SI-7).  

\textbf{Mitigate weaknesses in the systems and minimize the risk of hidden weakness}: Organizations should identify any latent weaknesses in their environment through penetration testing (CA-8), vulnerability scanning (RA-5), and system flaw remediation (SI-2). However, despite these measures, a system may have hidden weaknesses. Hence, organizations should validate any incoming data from external systems to prevent attacks related to insecure input such as cross-site scripting, SQL injection, and shell code execution (SI-10). Organizations should also deploy anti-malware services to prevent further malicious code executions (SI-3). 

\textbf{Observe system and control behavior}: Organizations should actively monitor every activity of system components (SI-4), and access to the system through system boundaries (SC-7) to identify any suspicious motives of adversaries. Finally, organizations should prepare criteria and metrics for assessing the performance of the enforced controls and consistently measure their performance (CA-7). 

\begin{table*}[]
    \centering
    \scriptsize
    \caption{Top 20 critical controls}
    \label{tab:top20}
    \begin{tabular}{p{3cm}lp{12cm}}
    \toprule
    
    \textbf{Control} & \textbf{Category} & \textbf{Description} \\ \midrule
    
    AC-2: Account Management & AC & Define the types of account permitted to access the system along with their roles and privileges \\
    
    AC-3: Access Enforcement & AC & Apply appropriate authorization to accounts during gaining an access to the system \\
    
    AC-4: Information Flow Enforcement & AC & Apply appropriate authorization to control information flow among the system components and with external systems \\
    
    AC-5: Separation of Duties & AC & Define different responsibilities for different accounts, or groups of accounts so that responsibilities do not overlap. E.g., individuals who have an administration role for creating new accounts for users should not have any role regarding auditing.   \\
    
    AC-6: Least Privilege & AC & Grant account with the minimum required privilege to perform the corresponding duties associated with the account role \\
    
    CA-7: Continuous Monitoring & CA & Determine metrics for assessing the effectiveness of the enforced controls and monitor the controls according to the metrics  \\
    
    CA-8: Penetration Testing & CA & Identify the weaknesses of the system from the attacker's point of view \\
    
    CM-2: Baseline Configuration & CM & Create a baseline configuration for the system components and periodically review, and update the configuration \\
    
    CM-5: Access Restrictions for Change & CM & In case of change in system components, restrict the access to the systems and permit only the authorized users to conduct the change of components \\
    
    CM-6: Configuration Settings & CM & Establish a secure configuration of the components according to the common secure configuration and hardening guidelines suggested by vendors, manufacturers, and federal agencies \\
    
    CM-7: Least Functionality & CM & Configure system components so that they can only perform the functions they need. For example, if an applications' task is to communicate with the \textit{HTTP} port, do not allow the application to access the \textit{SMTP} port  \\
    
    CM-8: System Component Inventory & CM & Document the number and types of every system component with unique identifiers with an appropriate level of granularity.   \\
    
    IA-2: Identification and Authentication & IA & Identify and authenticate organizational users while gaining access or performing their roles  \\
    
    RA-5: Vulnerability Monitoring and Scanning & RA & Continuously scan and monitor the presence of a vulnerability in the system components  \\
    
    SC-7: Boundary Protection & SC & Enforce and monitor access to the system components on the periphery of the systems such as routers, gateways, firewalls, remote-facing applications \\

    SI-2: Flaw Remediation & SI & Identify and remediate flaws among the system components \\ 
    
    SI-3: Malicious Code Protection & SI & Deploy anti-malware systems to block the execution of malicious code  \\

    SI-4: System Monitoring & SI & Monitor all the system components activities such as read, write, execution, and transfer operations \\
    
    SI-7: Software, Firmware, and Information Integrity & SI & Detect unauthorized changes made to system components such as software, firmware, databases  \\
    
    SI-10: Information Input Validation & SI & Apply sanity checks on information coming from external systems \\ \midrule
    
    \multicolumn{3}{p{17cm}}{AC: access control, CA: assessment, authorization, and monitoring, CM: configuration management, IA: identification and authorization, RA: risk assessment, SC: systems and communication protection, SI: systems and information integrity} \\
    
    \bottomrule
    \end{tabular}
    
\end{table*}

\section{Summary of the findings} \label{summary}
We summarize our findings from~\cref{sec:rq1}, and~\cref{sec:rq2} below. The evaluation of all 101 mitigating controls is reported in Table~\ref{tab:listofallcontrols} at the Appendix.
\begin{itemize}
    \item[\ding{43}] \textbf{The foremost four controls:} Organizations must implement the following four controls with the topmost prioritization: \textit{CM-2: Baseline configuration}, \textit{CM-6: Configuration settings}, \textit{SI-3: Malicious code protection}, and \textit{SI-4: System monitoring}. These four controls mitigate six of the top ten most risky techniques, shown in Table~\ref{tab:technique-risk}. However, the remaining four techniques do not have any control for mitigation;
    
    \item[\ding{43}] \textbf{The unmitigated but frequent adversary techniques:} New controls should be developed for gaining protections against \textit{TA0007: Discovery}-related techniques. Seven out of the top ten most frequent but unmitigated techniques among adversaries are from the \textit{Discovery} tactic;
    
    \item[\ding{43}] \textbf{The super eight controls:} We identify the four controls from the foremost four plus \textit{AC-3; Access enforcement}, \textit{AC-6: Least privilege}, \textit{CA-7: Continuous monitoring}, and \textit{CM-7: Least functionality} are the eight controls that appears top in all metrics except CR. However, these eight controls have relatively lower CR value but higher TEC. Overall, organizations can prioritize these controls as they come on top in every metric except CR. 
    
    \item[\ding{43}] \textbf{The top 20 critical controls:} Through applying clustering, we identify 20 controls shown in Table~\ref{tab:top20}, which, in combination, can mitigate 134 out of 135 mitigable techniques. These controls can mitigate, on average, 72\% of the techniques of 98\% of all cataloged adversary entities in ATT\&CK. These 20 controls contains all the control from \textit{super eight}. The controls primarily recommend organizations to secure access to the systems, apply the right configurations, mitigate weakness, and observe the behavior of systems and controls.  
    
\end{itemize}

\section{Takeaways}
\label{takeaways}
In this section, we discuss several takeaways from the metrics introduced in this study. 

\textbf{Technique Coverage (TEC) is the most important metric in the proposed metric suite.} We propose seven metrics in the metric suite, however, except in the case of CR, the top eight controls by the five metric: TAC, MTAC, AC, ATC, and CMR are identical to the top eight controls by the TEC. The reason is that the computation of TAC, MTAC, AC, ATC, and CMR metrics are based upon the mitigation of techniques in the respective tactics, and the corresponding adversaries using the techniques. Computing CMR requires the sum of risk mitigated by the techniques mitigated. Overall, the study provides seven metrics to analyse the control's mitigation ability, however, TEC is the most important metric of the seven. TEC not only provides a numeric measure on a control's mitigation coverage over techniques but also reflects how important the control is across a range of adversary techniques. For example, \textit{SI-4: System monitoring} mitigates 120 controls highlighting the substantial importance of monitoring everything in an environment if organizations need to defend themselves against a large number of techniques. One of the mitigated techniques by SI-4 is \textit{T1059: Command and scripting interpreter}. However, if SI-4 is implemented in desktops but not in mobile devices used in the organization, then an adversary can execute code in mobile devices. Thus, TEC provides a baseline to organization about a control's expected mitigation ability and trace any weaknesses in the implementation of controls.  

\textbf{Tactic Coverage (TAC) and Mean Tactic Coverage (MTAC) aid organizations in choosing controls against particular adversary tactics.} Organizations can identify what facet of their system is vulnerable to what particular tactics and then based upon TAC, they can prioritize what controls they need to implement in what facet of their systems. MTAC, on the other hand, indicates what control should be applied across all components because, adversaries can achieve their tactical goals on different components in the perimeter and the core components of the systems

\textbf{Control Redundancy (CR) emphasizes on mitigation diversity and reliability of a control's implementation. }
CR of a control also implies that how many diverse ways techniques can be mitigated, which captures a tradeoff situation for an organization. A relatively high CR means the corresponding technique can be mitigated in many ways but organizations have to implement multiple alternative controls. On the other hand, a relatively low CR value means organizations do not need to implement many alternative controls. However, they must implement the controls with low CR and they have to ensure the reliability of the control implementation, because, if the control fails, relatively low number of alternative exists.

\textbf{Adversarial Coverage (AC) and Adversarial Technique Coverage (ATC) aids organization in choosing controls against the changing threat landscape.} AC of a control implies how effective the control is against current trend of threat landscape. Based on AC, organizations can identify the need for new controls or re-assessment of existing controls. Organizations can collect data on their environment to identify what are the important controls for them based on their potential adversary entities. For example, several cybercrime groups only targets energy sectors, such as~\cite{hexane} while the others target only financial sectors, such as~\cite{admin338}. Thus, organizations can determine the set of important controls based on the relevant cybercrime groups in their corresponding business domains. ATC can aid organizations in determining what fraction of techniques for their respective cybercrime groups can be mitigated. If an organization identify that their controls are mitigating a low fraction of techniques from adversaries, then they should enhance or introduce new controls. They can also identify for what techniques, there is no mitigation, and for those cases, they can train employees and focus on more detection and recovering measures. For example, \textit{discovery} tactics have a relatively low number of mitigating controls. Thus, if an organization identifies a suspicious process trying to gather system information, they must shut down respective operations of those components to minimize attack surface. 

\textbf{Although, the Control's Mitigated Risk (CMR) is proportionate to technique coverage, the measure is data driven, and hence, CMR could vary with different organizations:} Our proposed risk metric quantifies the severity of techniques in terms of how they co-occur with other tactics based on the assumption that the more a technique co-occurs with other tactics, the more severe the technique is. We approximate the severity from the cataloged adversaries in ATT\&CK, which reflects cyberattacks on various domains of businesses. However, CMR could be calculated based on the data of co-occurring techniques obtained from a particular organization or environment. An organization can individually identify key assets and components in their system and find out what techniques affects them and then calculate the risk of a particular technique on a particular assets and then calculate the top controls in terms of mitigating risks on the asset or system facets

\textbf{Clustering the controls by the metrics aids organization in approximating the set of critical controls.} In this study we find two clusters, and based on the comparison shown in Table~\ref{tab:cluster-comparison}, organizations can choose \textit{Cluster-A} over \textit{Cluster-B}. We also identify that 13 out of top 20 critical controls~\cite{cisNetWrix} proposed by CIS controls~\cite{cistop20} has overlap with our identified top 20 controls. However, in case of future updates in the NIST controls, ATT\&CK techniques, and the mapping, the outcome of the cluster could be different. However, through clustering, an organization, who does not have any existing control, can obtain an approximate understanding on what group of controls have the most desired effectiveness in each metric. 

\textbf{The metric suite can aid in assessment of controls.} Organizations should monitor and identify adversarial activities inflicted upon in their systems and test if certain controls are performing as expected. Hence, our metrics can be used by an organization on their collected data to identify: (a) what controls are not working; (b) where controls should have been enforced; and (c) where to implement new controls. NIST provides a guideline~\cite{NIST80053A, NIST8011} where they suggests to examine, interview, and test all the controls enforced in an organization for continuous assessment. Our metrics can aid them in the examining phases of the controls based upon their ability to mitigate techniques. For example, one of the mapped techniques to \textit{SI-4: System monitoring} is \textit{T1204: User execution}. Hence, an organization can examine, in what capacity, their implemented system monitoring could aid them in mitigating user execution of malicious command. One way to mitigate user execution through monitoring is to checking the event log of all executed commands. If the organization identifies their monitoring activities do not cover checking execution logs, they can update the control accordingly.  

\section{Threats to validity}\label{limitation}
The metrics introduced in this study are based on the mapping (\cref{sec:concept-mapping}) and the ATT\&CK Version 10 TTPs. We evaluate the controls on the metrics, although the mapping cannot imply the whether an adversarial technique can fully or partially be mitigated by the mapped control(s). We also did not investigate the \textit{why} aspects of control mitigating techniques. For example, why \textit{SI-4} comes to the top in six metrics out of the seven. Moreover, frequently co-occurring techniques might have underlying factors that lead multiple controls mitigating the same set of techniques, cybercrime groups or malware. We advocate for further investigation on \textit{how} controls mitigate techniques, and what underlying factors are prevalent in mitigation. 

The proposed metric suite does not measure several constraints such as cost, implementation times, required resources, logistics and pedagogical support. Although, cost is a very important issue for choosing controls, in this study, we primarily aim to aware organizations about what security controls are the most important in terms of their mitigation ability over a set of documented adversarial techniques, and relevance against reported cybercrime groups and malware. As our metric suites provides a scalar value for each of the NIST controls, the metric suite can be incorporated with existing optimal control selection methods. The calculation of risk is based on the observed occurrence of techniques cited in ATT\&CK only. However, one dataset may not reflect the actual frequency of adversarial techniques. 

Overall, the metrics provide an initial baseline on decision making - what controls organizations need to identify, select, implement, and prioritize. Hence, organizations need to formulate additional metrics and criteria to measure how effectively the controls are working. We make the dataset and source code for this research made available for replication, including re-running the research when the ATT\&CK Framework and/or the ATT\&CK-NIST mapping are updated. We only investigate NIST SP800-53 controls and ATT\&CK techniques. However, organizations also use other control standards, such as ISO/IEC 27001~\cite{ISO27001} and CIS top 20~\cite{cistop20}. We encourage investigation on other control standards as well as future versions of ATT\&CK and NIST SP800-53 to gain a better understanding on the generalizable facts from our findings. 

\section{Related work}
\label{sec:related-work}
We discuss several related work in this section. Dutta et al.~\cite{dutta2019and} mapped the CIS top 20 controls to cyber-kill-chain phases~\cite{killchain}, NIST cybersecurity framework, vulnerabilities and assets. The authors then proposed an optimization technique to select the most cost effective security controls associated with the kill-chain phase probabilities, asset vulnerabilities and risk. Hadar et al.~\cite{hadarCyberDigitalTwin2020} proposed a simulator for automatic prioritization of security controls based on an attack graph of assets, vulnerabilities, and adversary attempts on the network devices. Their proposed framework identifies the missing controls on assets, and thus, minimize the attack surface, and they demonstrated their model on ISO/IEC 270001 controls. Kuppa et al.~\cite{kuppa2021linking} applied machine learning techniques to automatically map common vulnerabilities and exposure (CVE) to ATT\&CK techniques. They mapped 62,000 CVE records from past 10 years to 37 different ATT\&CK techniques. Researchers have investigated the selection and prioritization of control as well. Li et al.~\cite{liDefendingAdvancedPersistent2018} and Sawik et al.~\cite{sawikSelectionOptimalCountermeasure2013} proposed that optimal selection of control having multiple objectives from stakeholders such as mitigated risk, and cost. They found the optimal set based on epidemic models of malware spreading and integer programming respectively. Khajouei et al.~\cite{khajouei2017ranking}, and Imran et al.~\cite{tariq2020prioritization}. prioritized ISO/IEC 27001 controls through fuzzy analytic hierarchy process (AHP) technique. Al-Safwani et al.~\cite{al2018iscp} proposes a control prioritization model on TOPSIS technique. Overall, we find little research on the investigation of security controls mitigating adversarial attack techniques, and we investigate and evaluate the controls in terms of mitigation of adversarial techniques.

\section{Conclusion}
\label{sec:conclusion}
Organizations implement and enforce security controls by satisfying security requirements to avoid, detect, and mitigate the cyberthreats from adversaries. As implementing all available controls can be infeasible, organizations need to make an informed choice on the controls. To this end, we propose a security control metric suite to aid organizations in objectively identifying what controls they need to choose. We evaluate the NIST SP800-53 controls based on the metrics on adversarial techniques and tactics cataloged in MITRE ATT\&CK. We also evaluate the controls on adversarial techniques used in real-world cyberattacks by 125 cybercrime groups, and 544 malware cataloged in ATT\&CK. We identify a set of 20 critical controls that can mitigate 134 adversarial techniques, and on average, can mitigate 72\% of all techniques used by 98\% of the cataloged adversaries in MITRE ATT\&CK. We urge organizations, that do not have any controls enforced in place, to implement the top controls identified in the study. Findings of our paper can benefit cybersecurity practitioners and researchers by providing on organizational constraint agnostic (i.e. cost, risk, budget, logistics) baseline for decision making on the selection of controls. Moreover, organizations can formulate additional metrics and criteria to assess how effectively the controls are performing in their own organizations. 



\section*{Acknowledgment}
This work is supported by \textit{anonymous} funding.


\bibliographystyle{IEEEtran}
\bibliography{main}

\begin{thebibliography}{10}
\providecommand{\url}[1]{#1}
\csname url@samestyle\endcsname
\providecommand{\newblock}{\relax}
\providecommand{\bibinfo}[2]{#2}
\providecommand{\BIBentrySTDinterwordspacing}{\spaceskip=0pt\relax}
\providecommand{\BIBentryALTinterwordstretchfactor}{4}
\providecommand{\BIBentryALTinterwordspacing}{\spaceskip=\fontdimen2\font plus
\BIBentryALTinterwordstretchfactor\fontdimen3\font minus
  \fontdimen4\font\relax}
\providecommand{\BIBforeignlanguage}[2]{{%
\expandafter\ifx\csname l@#1\endcsname\relax
\typeout{** WARNING: IEEEtran.bst: No hyphenation pattern has been}%
\typeout{** loaded for the language `#1'. Using the pattern for}%
\typeout{** the default language instead.}%
\else
\language=\csname l@#1\endcsname
\fi
#2}}
\providecommand{\BIBdecl}{\relax}
\BIBdecl

\bibitem{hackernews-financial-backup}
S.~Khandelwal, ``{New Group of Hackers Targeting Businesses with Financially
  Motivated Cyber Attacks},''
  \url{https://thehackernews.com/2019/11/financial-cyberattacks.html}, 2019,
  [Online: accessed 10-Aug-2022].

\bibitem{reuters}
A.~Moon, ``{State-sponsored cyberattacks on banks on the rise: report},''
  \url{https://www.reuters.com/article/us-cyber-banks/state-sponsored-cyberattacks-on-banks-on-the-rise-report-idUSKCN1R32NJl},
  2019, [Online: accessed 10-Aug-2022].

\bibitem{checkpoint}
C.~Research, ``Cyberattack increased 50\% year over year,''
  \url{https://blog.checkpoint.com/2022/01/10/check-point-research-cyber-attacks-increased-50-year-over-year/}
  Accessed 10-Aug-2022, 2022.

\bibitem{forbes}
C.~Brooks, ``{Cybersecurity in 2022 A Fresh Look at Some Very Alarming
  Stats},''
  \url{https://www.forbes.com/sites/chuckbrooks/2022/01/21/cybersecurity-in-2022--a-fresh-look-at-some-very-alarming-stats/?sh=3aaf3c9e6b61},
  2022, [Online: accessed 15-Feb-2022].

\bibitem{betanews}
I.~Barker, ``{Cybercriminals can penetrate 93 percent of company networks},''
  \url{https://betanews.com/2021/12/20/cybercriminals-penetrate-93-percent-of-company-networks/},
  2022, [Online: accessed 15-Feb-2022].

\bibitem{control-definition}
``{NIST Glossary},''
  \url{https://csrc.nist.gov/glossary/term/security{\_}control}, 2022, [Online:
  accessed 10-Aug-2022].

\bibitem{NIST800r53}
NIST, ``Security and privacy controls for information systems and
  organizations,''
  \url{https://csrc.nist.gov/publications/detail/sp/800-53/rev-5/final}, 2022,
  [Online: Accessed 10-Aug-2022].

\bibitem{ttp-t1110}
``{Brute-force technique T1110 - Enterprise | MITRE ATT\&CK},''
  \url{https://attack.mitre.org/techniques/T1110/}, 2022, [Online: accessed
  10-Aug-2022].

\bibitem{ISO27001}
``{ISO/IEC 27001 Information Security Management},''
  \url{https://www.iso.org/isoiec-27001-information-security.html}, 2022,
  [Online: accessed 15-Feb-2022].

\bibitem{cistop20}
``{CIS Security Controls},'' \url{https://www.cisecurity.org/controls}, 2022,
  [Online: accessed 15-Feb-2022].

\bibitem{colonial-pipeline}
W.~Turton and K.~Mehrotra, ``Hackers breached colonial pipeline using
  compromised password,''
  \url{https://www.bloomberg.com/news/articles/2021-06-04/hackers-breached-colonial-pipeline-using-compromised-password}
  Accessed 10-Aug-2022, 2021.

\bibitem{mandiant-colonial-pipeline}
``{Mandiant: Compromised Colonial Pipeline password was reused},''
  \url{https://www.techtarget.com/searchsecurity/news/252502216/Mandiant-Compromised-Colonial-Pipeline-password-was-reused},
  2022, [Online: accessed: 10-Aug-2022].

\bibitem{wa-post}
E.~Nakashim and L.~Aratani, ``{DHS to issue first cybersecurity regulations for
  pipelines after Colonial hack},''
  \url{https://www.washingtonpost.com/business/2021/05/25/colonial-hack-pipeline-dhs-cybersecurity/},
  2021, [Online: accessed 10-Aug-2022].

\bibitem{us-gao}
U.~G.~A. Office, ``{Critical Infrastructure Protection: Agencies Need to Assess
  Adoption of Cybersecurity Guidance},''
  \url{https://www.gao.gov/products/gao-22-105103}, 2022, [Online: accessed
  10-Aug-2022].

\bibitem{bullguard-research}
P.~Lipman, ``One in five smbs don't use any cybersecurity,''
  \url{https://www.forbes.com/sites/forbestechcouncil/2020/03/02/one-in-five-smbs-dont-use-any-cybersecurity-heres-what-theyre-putting-at-risk/?sh=769422b47b95}
  Accessed 10-Aug-2022, 2020.

\bibitem{cnbc-research}
L.~Wronski, ``America’s small businesses aren’t ready for a cyberattack,''
  \url{https://www.cnbc.com/2022/05/21/americas-small-businesses-arent-ready-for-a-cyberattack.html}
  Accessed 10-Aug-2022, 2022.

\bibitem{hadarCyberDigitalTwin2020}
E.~Hadar, D.~Kravchenko, and A.~Basovskiy, ``Cyber {{Digital Twin Simulator}}
  for {{Automatic Gathering}} and {{Prioritization}} of {{Security Controls}}'
  {{Requirements}},'' in \emph{2020 {{IEEE}} 28th {{International Requirements
  Engineering Conference}} ({{RE}})}.\hskip 1em plus 0.5em minus 0.4em\relax
  {Zurich, Switzerland}: {IEEE}, Aug. 2020, pp. 250--259.

\bibitem{dutta2019and}
A.~Dutta and E.~Al-Shaer, ``“what”,“where”, and “why” cybersecurity
  controls to enforce for optimal risk mitigation,'' in \emph{2019 IEEE
  Conference on Communications and Network Security (CNS)}.\hskip 1em plus
  0.5em minus 0.4em\relax IEEE, 2019, pp. 160--168.

\bibitem{liDefendingAdvancedPersistent2018}
P.~Li, X.~Yang, Q.~Xiong, J.~Wen, and Y.~Y. Tang, ``Defending against the
  {{Advanced Persistent Threat}}: {{An Optimal Control Approach}},''
  \emph{Security and Communication Networks}, vol. 2018, pp. 1--14, 2018.

\bibitem{sawikSelectionOptimalCountermeasure2013}
T.~Sawik, ``Selection of optimal countermeasure portfolio in {{IT}} security
  planning,'' \emph{Decision Support Systems}, vol.~55, no.~1, pp. 156--164,
  Apr. 2013.

\bibitem{attack}
MITRE, ``{MITRE ATT\&CK},'' \url{https://attack.mitre.org/}, 2021, [Online:
  accessed 10-Aug-2022].

\bibitem{attack-control-mapping}
``Att\&ck control framework mappings,''
  \url{https://github.com/center-for-threat-informed-defense/attack-control-framework-mappings},
  2022, [Online: Accessed 10-Aug-2022].

\bibitem{CTID}
``Center for threat-informed defense,''
  \url{https://ctid.mitre-engenuity.org/}, 2022, [Online: Accessed
  10-Aug-2022].

\bibitem{gitrepo}
\url{https://github.com/brokenquark/nist-control-evaluation/tree/main}, 2022,
  [Online: Accessed 19-Aug-2022].

\bibitem{attack-design}
B.~Strom, A.~Applebaum, D.~Miller, K.~Nickels, A.~Pennington, and C.~Thomas,
  ``Mitre att\&ck: Design and philosophy,''
  \url{https://www.mitre.org/publications/technical-papers/mitre-attack-design-and-philosophy},
  MITRE, Tech. Rep., 2020.

\bibitem{nist-glossary}
``{NIST Glossary},''
  \url{https://csrc.nist.gov/glossary/term/Tactics_Techniques_and_Procedures},
  2022, [Online: accessed 10-Aug-2022].

\bibitem{ta0004}
``{Privilege Escalation Tactic TA0004 - Enterprise | MITRE ATT\&CK},''
  \url{https://attack.mitre.org/tactics/TA0004/}, 2022, [Online: accessed
  10-Aug-2022].

\bibitem{metasploit}
B.~Mckeague, ``{Pick-Six: Intercepting a FIN6 Intrusion, an Actor Recently Tied
  to Ryuk and LockerGoga Ransomware},''
  \url{https://www.gao.gov/products/gao-22-105103}, 2019, [Online: accessed
  10-Aug-2022].

\bibitem{fin6}
MITRE, ``{FIN6: Group G0037 | MITRE ATT\&CK},''
  \url{https://attack.mitre.org/groups/G0037/}, [Online: accessed 10-Aug-2022].

\bibitem{mitre-org}
``{The MITRE Corporation},'' \url{https://www.mitre.org/}, 2022, [Online:
  accessed 10-Aug-2022].

\bibitem{attack-groups-def}
``{Groups - Enterprise | MITRE ATT\&CK},''
  \url{https://attack.mitre.org/groups/}, 2022, [Online: accessed 10-Aug-2022].

\bibitem{attack-malware-def}
``{Software - Enterprise | MITRE ATT\&CK},''
  \url{https://attack.mitre.org/software/}, 2022, [Online: accessed
  10-Aug-2022].

\bibitem{engenuity}
``Mitre engenuity,'' \url{https://mitre-engenuity.org/}, 2022, [Online:
  Accessed 10-Aug-2022].

\bibitem{T1185}
``{Browser session hijacking T1185 - Enterprise | MITRE ATT\&CK},''
  \url{https://attack.mitre.org/techniques/T1185/}, 2022, [Online: accessed
  10-Aug-2022].

\bibitem{pendletonSurveySystemsSecurity2017}
M.~Pendleton, R.~{Garcia-Lebron}, J.-H. Cho, and S.~Xu, ``A {{Survey}} on
  {{Systems Security Metrics}},'' \emph{ACM Computing Surveys}, vol.~49, no.~4,
  pp. 1--35, Dec. 2017.

\bibitem{hubbardHowMeasureAnything2016}
D.~W. Hubbard and R.~Seiersen, \emph{How to {{Measure Anything}} in
  {{Cybersecurity Risk}}}, 1st~ed.\hskip 1em plus 0.5em minus 0.4em\relax
  {Wiley}, Aug. 2016.

\bibitem{admin338}
``{admin@338, Group G0018 | MITRE ATT\&CK},''
  \url{https://attack.mitre.org/groups/G0018/}, [Online: accessed 10-Aug-2022].

\bibitem{T1059}
``{Command and Scripting Interpreter Technique T1059 - Enterprise | MITRE
  ATT\&CK},'' \url{https://attack.mitre.org/techniques/T1059/}, 2022, [Online:
  accessed 10-Aug-2022].

\bibitem{T1566}
``{Phishing T1566 - Enterprise | MITRE ATT\&CK},''
  \url{https://attack.mitre.org/techniques/T1566/}, 2022, [Online: accessed
  10-Aug-2022].

\bibitem{T1204}
``{User Execution T1204 - Enterprise | MITRE ATT\&CK},''
  \url{https://attack.mitre.org/techniques/T1204/}, 2022, [Online: accessed
  10-Aug-2022].

\bibitem{aptC36}
``{APT-C-36, Group G0009 | MITRE ATT\&CK},''
  \url{https://attack.mitre.org/groups/G0099/}, [Online: accessed 10-Aug-2022].

\bibitem{T1027}
``{Obfuscated Files and Information Technique T1027 - Enterprise | MITRE
  ATT\&CK},'' \url{https://attack.mitre.org/techniques/T1027/}, 2022, [Online:
  accessed 10-Aug-2022].

\bibitem{T1105}
``{Ingress Tool Transfer Technique T1105 - Enterprise | MITRE ATT\&CK},''
  \url{https://attack.mitre.org/techniques/T1105/}, 2022, [Online: accessed
  10-Aug-2022].

\bibitem{TA0002}
``{Execution Tactic TA0002 - Enterprise | MITRE ATT\&CK},''
  \url{https://attack.mitre.org/tactics/TA0002/}, 2022, [Online: accessed
  10-Aug-2022].

\bibitem{TA0011}
``{Command and Control Tactic TA0011 - Enterprise | MITRE ATT\&CK},''
  \url{https://attack.mitre.org/tactics/TA0011/}, 2022, [Online: accessed
  10-Aug-2022].

\bibitem{TA0001}
``{Initial Access Tactic TA0001 - Enterprise | MITRE ATT\&CK},''
  \url{https://attack.mitre.org/tactics/TA0001/}, 2022, [Online: accessed
  10-Aug-2022].

\bibitem{working-with-attack}
``{Working with ATT\&CK | MITRE},''
  \url{https://attack.mitre.org/resources/working-with-attack/}, 2022, [Online:
  accessed 10-Aug-2022].

\bibitem{dbattacktactics}
``Enterprise att\&ck tactics,''
  \url{https://attack.mitre.org/docs/enterprise-attack-v10.1/enterprise-attack-v10.1-tactics.xlsx},
  2022, [Online: Accessed 10-Aug-2022].

\bibitem{db-attack-techniques}
``Enterprise att\&ck techniques,''
  \url{https://attack.mitre.org/docs/enterprise-attack-v10.1/enterprise-attack-v10.1-techniques.xlsx},
  2022, [Online: Accessed 10-Aug-2022].

\bibitem{db-attack-groups}
``Enterprise att\&ck groups,''
  \url{https://attack.mitre.org/docs/enterprise-attack-v10.1/enterprise-attack-v10.1-groups.xlsx},
  2022, [Online: Accessed 10-Aug-2022].

\bibitem{db-attack-malware}
``Enterprise att\&ck software,''
  \url{https://attack.mitre.org/docs/enterprise-attack-v10.1/enterprise-attack-v10.1-software.xlsx},
  2022, [Online: Accessed 10-Aug-2022].

\bibitem{abdi2010principal}
H.~Abdi and L.~J. Williams, ``Principal component analysis,'' \emph{Wiley
  interdisciplinary reviews: computational statistics}, vol.~2, no.~4, pp.
  433--459, 2010.

\bibitem{thorndike1953belongs}
R.~L. Thorndike, ``Who belongs in the family,'' in \emph{Psychometrika}.\hskip
  1em plus 0.5em minus 0.4em\relax Citeseer, 1953.

\bibitem{Lloyd1982Least}
S.~Lloyd, ``Least squares quantization in pcm,'' \emph{IEEE Transactions on
  Information Theory}, vol.~28, no.~2, pp. 129--137, 1982.

\bibitem{TA0003}
``{Persistence Tactic TA0003 - Enterprise | MITRE ATT\&CK},''
  \url{https://attack.mitre.org/tactics/TA0003/}, 2022, [Online: accessed
  10-Aug-2022].

\bibitem{TA0005}
``{Defense Evasion Tactic TA0005 - Enterprise | MITRE ATT\&CK},''
  \url{https://attack.mitre.org/tactics/TA0005/}, 2022, [Online: accessed
  10-Aug-2022].

\bibitem{TA0006}
``{Credential Access Tactic TA0006 - Enterprise | MITRE ATT\&CK},''
  \url{https://attack.mitre.org/tactics/TA0006/}, 2022, [Online: accessed
  10-Aug-2022].

\bibitem{TA0007}
``{Discovery Tactic TA0007 - Enterprise | MITRE ATT\&CK},''
  \url{https://attack.mitre.org/tactics/TA0007/}, 2022, [Online: accessed
  10-Aug-2022].

\bibitem{TA0008}
``{Lateral Movement Tactic TA0008 - Enterprise | MITRE ATT\&CK},''
  \url{https://attack.mitre.org/tactics/TA0008/}, 2022, [Online: accessed
  10-Aug-2022].

\bibitem{TA0009}
``{Collection Tactic TA0009 - Enterprise | MITRE ATT\&CK},''
  \url{https://attack.mitre.org/tactics/TA0009/}, 2022, [Online: accessed
  10-Aug-2022].

\bibitem{TA0010}
``{Exfiltration Tactic TA0010 - Enterprise | MITRE ATT\&CK},''
  \url{https://attack.mitre.org/tactics/TA0010/}, 2022, [Online: accessed
  10-Aug-2022].

\bibitem{TA0040}
``{Impact Tactic TA0040 - Enterprise | MITRE ATT\&CK},''
  \url{https://attack.mitre.org/tactics/TA0040/}, 2022, [Online: accessed
  10-Aug-2022].

\bibitem{TA0042}
``{Resource Development Tactic TA0042 - Enterprise | MITRE ATT\&CK},''
  \url{https://attack.mitre.org/tactics/TA0042/}, 2022, [Online: accessed
  10-Aug-2022].

\bibitem{TA0043}
``{Reconnaissance Tactic TA0043 - Enterprise | MITRE ATT\&CK},''
  \url{https://attack.mitre.org/tactics/TA0043/}, 2022, [Online: accessed
  10-Aug-2022].

\bibitem{mitigationM1056}
``{Pre-compromise M1056 - Enterprise | MITRE ATT\&CK},''
  \url{https://attack.mitre.org/mitigations/M1056/}, 2022, [Online: accessed
  10-Aug-2022].

\bibitem{T1082}
``{System Information Discovery T1082 - Enterprise | MITRE ATT\&CK},''
  \url{https://attack.mitre.org/techniques/T1082/}, 2022, [Online: accessed
  10-Aug-2022].

\bibitem{T1083}
``{File and directory discovery T1083 - Enterprise | MITRE ATT\&CK},''
  \url{https://attack.mitre.org/techniques/T1083/}, 2022, [Online: accessed
  10-Aug-2022].

\bibitem{T1057}
``{Process discovery T1057 - Enterprise | MITRE ATT\&CK},''
  \url{https://attack.mitre.org/techniques/T1057/}, 2022, [Online: accessed
  10-Aug-2022].

\bibitem{T1016}
``{System Network Configuration Discovery T1016 - Enterprise | MITRE
  ATT\&CK},'' \url{https://attack.mitre.org/techniques/T1016/}, 2022, [Online:
  accessed 10-Aug-2022].

\bibitem{T1140}
``{Deobfuscate/decode files or information T1140 - Enterprise | MITRE
  ATT\&CK},'' \url{https://attack.mitre.org/techniques/T1140/}, 2022, [Online:
  accessed 10-Aug-2022].

\bibitem{T1033}
``{System Owner/User discovery T1033 - Enterprise | MITRE ATT\&CK},''
  \url{https://attack.mitre.org/techniques/T1033/}, 2022, [Online: accessed
  10-Aug-2022].

\bibitem{T1113}
``{Screen capture T1113 - Enterprise | MITRE ATT\&CK},''
  \url{https://attack.mitre.org/techniques/T1113/}, 2022, [Online: accessed
  10-Aug-2022].

\bibitem{T1518}
``{Software discovery T1518 - Enterprise | MITRE ATT\&CK},''
  \url{https://attack.mitre.org/techniques/T1518/}, 2022, [Online: accessed
  10-Aug-2022].

\bibitem{T1074}
``{Data Staged T1074 - Enterprise | MITRE ATT\&CK},''
  \url{https://attack.mitre.org/techniques/T1074/}, 2022, [Online: accessed
  10-Aug-2022].

\bibitem{T1012}
``{Query Registry T1012 - Enterprise | MITRE ATT\&CK},''
  \url{https://attack.mitre.org/techniques/T1012/}, 2022, [Online: accessed
  10-Aug-2022].

\bibitem{T1071}
``{Application Layer Protocol Technique T1071 - Enterprise | MITRE ATT\&CK},''
  \url{https://attack.mitre.org/techniques/T1071/}, 2022, [Online: accessed
  10-Aug-2022].

\bibitem{T1070}
``{Indicator Removal on Host Technique T1070 - Enterprise | MITRE ATT\&CK},''
  \url{https://attack.mitre.org/techniques/T1070/}, 2022, [Online: accessed
  10-Aug-2022].

\bibitem{T1547}
``{Boot or Logon autostart execution T1547 - Enterprise | MITRE ATT\&CK},''
  \url{https://attack.mitre.org/techniques/T1547/}, 2022, [Online: accessed
  10-Aug-2022].

\bibitem{hexane}
``{Hexane, Group G1001 | MITRE ATT\&CK},''
  \url{https://attack.mitre.org/groups/G1001/}, [Online: accessed 10-Aug-2022].

\bibitem{cisNetWrix}
R.~Brooks, ``Top 20 critical security controls for effective cyber defense,''
  \url{https://blog.netwrix.com/2018/02/01/top-20-critical-security-controls-for-effective-cyber-defense/}
  Accessed 10-Aug-2022, 2022.

\bibitem{NIST80053A}
NIST, ``Assessing security and privacy controls in information systems and
  organizations,''
  \url{https://www.nist.gov/publications/assessing-security-and-privacy-controls-information-systems-and-organizations},
  2022, [Online: Accessed 10-Aug-2022].

\bibitem{NIST8011}
------, ``Automation support for security control assessments: Software
  vulnerability management--nist publishes nistir 8011 vol. 4,''
  \url{https://csrc.nist.gov/publications/detail/nistir/8011/vol-1/final},
  2022, [Online: Accessed 10-Aug-2022].

\bibitem{killchain}
L.~Martin, ``{Cyber Kill-Chain},''
  \url{https://www.lockheedmartin.com/en-us/capabilities/cyber/cyber-kill-chain.html},
  [Online: accessed 10-Aug-2022].

\bibitem{kuppa2021linking}
A.~Kuppa, L.~Aouad, and N.-A. Le-Khac, ``Linking cve’s to mitre att\&ck
  techniques,'' in \emph{The 16th International Conference on Availability,
  Reliability and Security}, 2021, pp. 1--12.

\bibitem{khajouei2017ranking}
H.~Khajouei, M.~Kazemi, and S.~H. Moosavirad, ``Ranking information security
  controls by using fuzzy analytic hierarchy process,'' \emph{Information
  Systems and e-Business Management}, vol.~15, no.~1, pp. 1--19, 2017.

\bibitem{tariq2020prioritization}
M.~I. Tariq, S.~Ahmed, N.~A. Memon, S.~Tayyaba, M.~W. Ashraf, M.~Nazir,
  A.~Hussain, V.~E. Balas, and M.~M. Balas, ``Prioritization of information
  security controls through fuzzy ahp for cloud computing networks and wireless
  sensor networks,'' \emph{Sensors}, vol.~20, no.~5, p. 1310, 2020.

\bibitem{al2018iscp}
N.~Al-Safwani, Y.~Fazea, and H.~Ibrahim, ``Iscp: In-depth model for selecting
  critical security controls,'' \emph{Computers \& Security}, vol.~77, pp.
  565--577, 2018.

\end{thebibliography}

\section*{Appendix}

\onecolumn

\begin{longtable}{lrrrrrr}
    \caption{Evaluation of 101 mitigating controls based on the proposed security control metric suite}
    \label{tab:listofallcontrols} \\ 
    
    \toprule
    
    \textbf{Control} & \textbf{TEC} & \textbf{MTAC} & \textbf{CR} & \textbf{AC} & \textbf{ATC} & \textbf{CMR} \\ \midrule
    
    SI-4 & 120 & 0.68 & 14.25 & 0.98 & 0.71 & 8.03 \\
    CM-6 & 111 & 0.63 & 14.49 & 0.98 & 0.66 & 7.37 \\
    CM-2 & 97 & 0.55 & 14.73 & 0.96 & 0.61 & 6.70 \\
    AC-3 & 91 & 0.55 & 15.59 & 0.94 & 0.47 & 5.31 \\
    CM-7 & 85 & 0.47 & 14.82 & 0.95 & 0.51 & 5.57 \\
    CA-7 & 84 & 0.51 & 15.65 & 0.96 & 0.57 & 6.32 \\
    AC-6 & 83 & 0.49 & 16.00 & 0.92 & 0.42 & 4.88 \\
    SI-3 & 81 & 0.46 & 14.19 & 0.96 & 0.62 & 6.96 \\
    SI-7 & 70 & 0.39 & 16.81 & 0.92 & 0.41 & 4.51 \\
    SC-7 & 68 & 0.39 & 14.32 & 0.88 & 0.33 & 3.68 \\
    AC-4 & 67 & 0.39 & 15.75 & 0.90 & 0.36 & 4.00 \\
    AC-2 & 66 & 0.39 & 17.44 & 0.90 & 0.37 & 4.26 \\
    IA-2 & 60 & 0.36 & 17.18 & 0.88 & 0.33 & 3.71 \\
    AC-5 & 53 & 0.32 & 18.25 & 0.88 & 0.32 & 3.57 \\
    CM-5 & 51 & 0.31 & 19.10 & 0.86 & 0.29 & 3.24 \\
    RA-5 & 46 & 0.29 & 18.50 & 0.84 & 0.24 & 2.81 \\
    CM-8 & 42 & 0.27 & 19.50 & 0.80 & 0.21 & 2.32 \\
    SI-10 & 36 & 0.19 & 16.19 & 0.87 & 0.25 & 2.76 \\
    SI-2 & 34 & 0.20 & 18.79 & 0.85 & 0.27 & 2.90 \\
    CA-8 & 34 & 0.20 & 19.68 & 0.80 & 0.18 & 2.04 \\
    AC-17 & 30 & 0.18 & 18.53 & 0.78 & 0.17 & 1.96 \\
    AC-16 & 26 & 0.15 & 19.96 & 0.71 & 0.15 & 1.66 \\
    AC-20 & 26 & 0.20 & 18.73 & 0.46 & 0.08 & 1.02 \\
    IA-5 & 23 & 0.15 & 18.48 & 0.35 & 0.07 & 0.78 \\
    SI-15 & 21 & 0.11 & 16.86 & 0.67 & 0.11 & 1.19 \\
    IA-4 & 16 & 0.10 & 23.00 & 0.61 & 0.10 & 1.14 \\
    SC-18 & 15 & 0.09 & 20.60 & 0.70 & 0.11 & 1.28 \\
    SA-11 & 15 & 0.09 & 21.53 & 0.51 & 0.07 & 0.77 \\
    SC-28 & 15 & 0.10 & 21.40 & 0.41 & 0.06 & 0.69 \\
    SI-12 & 14 & 0.08 & 21.07 & 0.50 & 0.07 & 0.77 \\
    SC-3 & 14 & 0.08 & 21.21 & 0.35 & 0.06 & 0.63 \\
    SA-8 & 13 & 0.10 & 18.92 & 0.57 & 0.08 & 0.98 \\
    CM-3 & 12 & 0.06 & 21.67 & 0.73 & 0.13 & 1.35 \\
    IA-8 & 12 & 0.08 & 21.83 & 0.71 & 0.12 & 1.19 \\
    SC-4 & 12 & 0.06 & 20.42 & 0.48 & 0.06 & 0.65 \\
    AC-19 & 12 & 0.07 & 21.08 & 0.44 & 0.05 & 0.59 \\
    CM-11 & 11 & 0.07 & 21.00 & 0.73 & 0.14 & 1.53 \\
    IA-9 & 11 & 0.05 & 17.36 & 0.70 & 0.12 & 1.26 \\
    CP-9 & 11 & 0.05 & 13.73 & 0.54 & 0.08 & 0.82 \\
    SA-10 & 11 & 0.07 & 22.27 & 0.53 & 0.07 & 0.77 \\
    SC-39 & 11 & 0.07 & 21.36 & 0.45 & 0.06 & 0.64 \\
    AC-18 & 11 & 0.06 & 20.27 & 0.42 & 0.04 & 0.52 \\
    AC-23 & 10 & 0.08 & 19.50 & 0.42 & 0.05 & 0.67 \\
    SC-23 & 8 & 0.04 & 16.00 & 0.57 & 0.07 & 0.79 \\
    CP-7 & 8 & 0.04 & 13.12 & 0.42 & 0.04 & 0.52 \\
    SC-8 & 8 & 0.05 & 19.88 & 0.29 & 0.03 & 0.37 \\
    SC-2 & 8 & 0.05 & 21.25 & 0.16 & 0.02 & 0.21 \\
    SC-34 & 7 & 0.04 & 19.57 & 0.26 & 0.03 & 0.30 \\
    SC-29 & 7 & 0.04 & 21.57 & 0.16 & 0.02 & 0.20 \\
    SC-30 & 7 & 0.04 & 21.57 & 0.16 & 0.02 & 0.20 \\
    AC-7 & 6 & 0.06 & 21.33 & 0.18 & 0.02 & 0.29 \\
    SA-15 & 6 & 0.05 & 25.33 & 0.20 & 0.02 & 0.26 \\
    CP-10 & 6 & 0.03 & 11.83 & 0.09 & 0.01 & 0.13 \\
    SC-20 & 5 & 0.03 & 13.80 & 0.54 & 0.06 & 0.69 \\
    SC-44 & 5 & 0.02 & 13.60 & 0.28 & 0.04 & 0.48 \\
    SI-16 & 5 & 0.03 & 18.40 & 0.38 & 0.04 & 0.47 \\
    SI-8 & 5 & 0.03 & 12.20 & 0.20 & 0.03 & 0.37 \\
    CM-10 & 5 & 0.03 & 21.20 & 0.26 & 0.03 & 0.32 \\
    SC-12 & 5 & 0.03 & 21.20 & 0.28 & 0.03 & 0.30 \\
    CP-2 & 5 & 0.03 & 9.60 & 0.08 & 0.01 & 0.12 \\
    IA-7 & 5 & 0.02 & 19.00 & 0.12 & 0.01 & 0.12 \\
    MP-7 & 5 & 0.04 & 10.00 & 0.06 & 0.01 & 0.07 \\
    SC-31 & 4 & 0.03 & 17.25 & 0.48 & 0.05 & 0.60 \\
    CP-6 & 4 & 0.02 & 16.75 & 0.41 & 0.04 & 0.45 \\
    SC-13 & 4 & 0.02 & 15.25 & 0.29 & 0.03 & 0.36 \\
    SA-17 & 4 & 0.04 & 17.25 & 0.21 & 0.02 & 0.24 \\
    CA-3 & 4 & 0.03 & 16.75 & 0.20 & 0.02 & 0.23 \\
    AC-12 & 4 & 0.03 & 21.75 & 0.13 & 0.01 & 0.13 \\
    IA-6 & 4 & 0.02 & 23.75 & 0.11 & 0.01 & 0.11 \\
    SC-26 & 4 & 0.02 & 23.75 & 0.06 & 0.00 & 0.07 \\
    SC-35 & 4 & 0.02 & 23.75 & 0.06 & 0.00 & 0.07 \\
    SI-5 & 4 & 0.02 & 23.75 & 0.06 & 0.00 & 0.07 \\
    SC-41 & 4 & 0.04 & 10.75 & 0.06 & 0.01 & 0.07 \\
    IA-3 & 4 & 0.02 & 27.25 & 0.06 & 0.00 & 0.05 \\
    SI-14 & 3 & 0.02 & 23.67 & 0.37 & 0.04 & 0.40 \\
    SC-36 & 3 & 0.01 & 19.00 & 0.38 & 0.03 & 0.40 \\
    SA-22 & 3 & 0.03 & 17.33 & 0.22 & 0.02 & 0.23 \\
    SA-4 & 3 & 0.04 & 20.67 & 0.20 & 0.02 & 0.22 \\
    SA-9 & 3 & 0.02 & 17.33 & 0.19 & 0.01 & 0.20 \\
    SA-3 & 3 & 0.03 & 26.00 & 0.17 & 0.02 & 0.19 \\
    CA-2 & 3 & 0.02 & 23.33 & 0.07 & 0.01 & 0.07 \\
    AC-10 & 3 & 0.01 & 14.67 & 0.03 & 0.00 & 0.03 \\
    SC-21 & 2 & 0.01 & 12.00 & 0.44 & 0.04 & 0.44 \\
    SC-22 & 2 & 0.01 & 12.00 & 0.44 & 0.04 & 0.44 \\
    SC-16 & 2 & 0.01 & 17.00 & 0.27 & 0.02 & 0.26 \\
    AC-21 & 2 & 0.02 & 26.00 & 0.19 & 0.02 & 0.19 \\
    SC-38 & 2 & 0.01 & 12.00 & 0.18 & 0.01 & 0.19 \\
    SA-16 & 2 & 0.03 & 24.00 & 0.17 & 0.01 & 0.17 \\
    AC-11 & 2 & 0.02 & 26.50 & 0.11 & 0.01 & 0.11 \\
    SC-17 & 2 & 0.02 & 13.50 & 0.01 & 0.00 & 0.01 \\
    SC-10 & 1 & 0.00 & 17.00 & 0.43 & 0.04 & 0.40 \\
    SC-37 & 1 & 0.00 & 17.00 & 0.43 & 0.04 & 0.40 \\
    AC-22 & 1 & 0.02 & 22.00 & 0.19 & 0.02 & 0.18 \\
    AC-24 & 1 & 0.02 & 22.00 & 0.19 & 0.02 & 0.18 \\
    AC-25 & 1 & 0.02 & 22.00 & 0.19 & 0.02 & 0.18 \\
    IR-5 & 1 & 0.00 & 20.00 & 0.12 & 0.01 & 0.12 \\
    SC-6 & 1 & 0.00 & 20.00 & 0.12 & 0.01 & 0.12 \\
    IA-11 & 1 & 0.00 & 13.00 & 0.06 & 0.00 & 0.05 \\
    AC-14 & 1 & 0.00 & 13.00 & 0.01 & 0.00 & 0.01 \\
    AC-8 & 1 & 0.01 & 6.00 & 0.01 & 0.00 & 0.01 \\
    MA-5 & 1 & 0.00 & 6.00 & 0.00 & 0.00 & 0.00 \\

    \bottomrule
    
\end{longtable}

\end{document}